\RequirePackage{amsmath}
\documentclass[runningheads]{llncs}
\usepackage{setspace}
\usepackage{color}
\usepackage[hidelinks]{hyperref}
\usepackage{graphicx}
\usepackage{tikz}
\usepackage{float}%
\usetikzlibrary{positioning,shapes.geometric,decorations.pathmorphing}%positioning,decorations.pathmorphing
\tikzset{every picture/.style={line width=1pt},
root/.style={ regular polygon, regular polygon sides=3, rotate=180,fill=black,inner sep=0.5mm},
tree/.style={circle,inner sep=0.6mm,fill=black},
leaf/.style={anchor=north,circle,draw,inner sep=0.6mm},
ret/.style={regular polygon, regular polygon sides=4,fill=black,inner sep=0.6mm},
subA/.style={draw,anchor=north,regular polygon, regular polygon sides=3,inner sep=1mm}}
\usepackage{mathtools}
\usepackage{algorithm}
\usepackage{algpseudocode}
\usepackage{amssymb} %to use nexist
\MakeRobust{\Call}
\newtheorem{cor}{Corollary}
\usepackage[normalem]{ulem}
\usepackage{thmtools, thm-restate}
\usepackage[titletoc]{appendix}
\let\oldforall\forall\renewcommand{\forall}{\:\oldforall\:}
\let\oldexists\exists\renewcommand{\exists}{\:\oldexists\:}
\newcommand{\N}{\mathcal{N}}

\newcommand{\B}{\mathcal{B}}
\newcommand{\cs}[2]{{#1}\langle{#2}\rangle}
\renewcommand{\above}{reach^-}
\newcommand{\below}{reach}
\newcommand{\crof}{\subseteq_{cr}}

\newcounter{commentcounter}
\definecolor{MLcommentcolor}{RGB}{50,50,220}
\definecolor{KLcommentcolor}{RGB}{250,50,50}
\definecolor{OTScommentcolor}{RGB}{0,250,120}
\newcommand{\comment}[2]{}
\algnewcommand\algorithmicforeach{\textbf{for each}}
\algdef{S}[FOR]{ForEach}[1]{\algorithmicforeach\ #1\ \algorithmicdo}

\newcommand{\macrs}{\textsc{MACRS}}
\newcommand{\macrssimple}{\textsc{MACRS-Simple}}

\newcommand{\rettrimmed}{\textsc{Reticulation-Trimmed Enumeration}}

\title{Finding agreement cherry-reduced subnetworks in level-1 networks}

\author{Kaari Landry*\inst{1}%\orcidID{0000-0003-3292-5633} 
\and
Olivier Tremblay-Savard\inst{1}%
\and
Manuel Lafond\inst{2}%\orcidID{2222--3333-4444-5555}
}
\authorrunning{K. Landry et al.}
\institute{
University of Manitoba, Winnipeg MB, Canada
\email{* landryk1@cs.umanitoba.ca}
\and
Université de Sherbrooke, Sherbrooke QC, Canada
}

\begin{document}
%\usetikzlibrary{shapes.geometric}
\maketitle
\vspace{-2em}
\begin{abstract}
    Phylogenetic networks are increasingly being considered as better suited to represent the complexity of the evolutionary relationships between species. One class of phylogenetic networks that has received a lot of attention recently is the class of orchard networks, which is composed of networks that can be reduced to a single leaf using cherry reductions. Cherry reductions, also called cherry-picking operations, remove either a leaf of a simple cherry (sibling leaves sharing a parent) or a reticulate edge of a reticulate cherry (two leaves whose parents are connected by a reticulate edge). In this paper, we present a fixed-parameter tractable algorithm to solve the problem of finding a maximum agreement cherry-reduced subnetwork (MACRS) between two rooted binary level-1 networks. This is first exact algorithm proposed to solve the MACRS problem. As proven in earlier work, there is a direct relationship between finding an MACRS and calculating a distance based on cherry operations. As a result, the proposed algorithm also provides a distance that can be used for the comparison of level-1 networks.

\end{abstract}

%\tableofcontents
% -----------------
\section{Introduction}
        %tex master: main

%\ml{There have been several new papers on cherry reductions and we need to make sure we list them all for a proper litt review.

%We also need to check the relationships between orchards and other network classes, to motivate them.  This can help addressing the criticism that our metric cannot compare any two networks.
%For example, all tree-child networks are orchards, and so our approach gives a metric on tree-child networks, which have been used in [XYZ].
%}

%\kl{orchards can be constructed from general networks. linz+semple "attaching leaves and picking cherries..."}

%\ml{[ML: so the title is so long that it is suppressed in the page headers :PI'd be ok with cutting a few keywords if they appear in the abstract, e.g. ``Finding agreement cherry-reduced subnetworks in level-1 networks'']}KL: i put this as the title.

%\ml{[ML: a general note, I added ``orchard'' in the problem statements, because none of this applies to networks in general.  If possible, it'd be good to skim through the paper with that in mind, in case things we say are false because we had orchards in mind (not crucial though).]}
%\kl{KL: end of 2.2 has statement indicating we assume networks are orchards}

Phylogenetic trees have been used extensively throughout the years to represent simple evolutionary relationships between species. Because of this, many tools and techniques are readily available to efficiently build, compare and evaluate trees. Phylogenetic networks on the other hand are much better suited to represent more complex relationships, such as the ones resulting from hybridization, recombination and lateral gene transfer events~\cite{10.1093/molbev/msj030}. In the last 15 years or so, bioinformatics research has focused increasingly on solving problems related to phylogenetic networks, such as network construction~\cite{park2010bootstrap,nguyen2015likelihood,solis2017phylonetworks,wen2018inferring,tan2019qs,allen2020estimating,lutteropp2022netrax}, minimum hybridization number~\cite{baroni2005framework,humphries2013cherry,van2019polynomial,janssen2020combining,huber2021rigid,huber2022cherry,bernardini2022reconstructing}, tree/network containment~\cite{janssen2020linear,janssen2021cherry,van2022embedding}, and distance calculation between networks~\cite{cardona2008metrics,lu2017program,CDIST22}.

One crucial concept that has been shown to be a very useful tool in solving several of the important phylogenetic network problems mentioned above is the one of cherry-picking sequences~\cite{humphries2013cherry,linz2019attaching}. A cherry-picking sequence is made up of operations that can reduce a network by either removing one leaf of a simple (tree-like) cherry ({\it i.e.} two leaf siblings descending from the same parent vertex), or removing one reticulate edge of a reticulated cherry (two leaves whose parent vertices are connected by a reticulate edge). The concept of cherry-picking has been so valuable that it led to the definition of {\it orchard networks}, also known as cherry-picking networks, which are simply phylogenetic networks that can be reduced to a single leaf by cherry-picking operations~\cite{erdHos2019class,janssen2021cherry}. Recent work has been focusing on further characterizing and classifying different subtypes of orchard networks~\cite{van2021unifying,kong2022classes,van2022orchard}.

Lately, we have used a generalized definition of {\it cherry operations} to describe both cherry reductions ({\it i.e}. cherry picking) and cherry expansions (the reverse of a reduction, which adds a simple or reticulate cherry)~\cite{CDIST22}. We have then defined four novel distances between orchard networks that are based on cherry operations, with three of them being different formulations of an equivalent distance (construction, deconstruction and tail distances) and the fourth one (mixed distance) being a lower bound for the other three. In the process of describing these distances, the concept of a {\it maximum agreement cherry-reduced subnetwork} (MACRS -- note that we replace cherry-picking used in~\cite{CDIST22} by cherry-reduced here for clarity) was defined to represent a network contained in both networks being compared that maximizes the number of vertices. We showed that finding an MACRS of two orchard networks was NP-hard, and this was analogous to the problem of calculating the three equivalent distances.

In this work, we present an exact fixed-parameter tractable (FPT) algorithm to compute an MACRS of two rooted binary level-1 networks that is exponential in the sum of reticulations present in both networks. More precisely, our algorithm runs in $O(3^r n^3)$, where $r$ is the sum of reticulations and n represents the maximum number of vertices of the input networks. Our approach essentially consists of enumerating a certain set of subnetworks of the input networks in which all possible combinations of reticulation edges have been removed. 
Then, it makes use of a dynamic programming algorithm that finds whether there is an MACRS (and what it is, if it exists) or not between two level-l networks in which reticulations that are remaining cannot be removed (we call this problem {\it MACRS-Simple}). We prove that the initial MACRS problem can be solved by solving the MACRS-Simple problem on all combinations of enumerated subnetworks. 

It is worth noting another important difference between the previous defining work on MACRS and this article is the definition of networks. Specifically, we allow leaves of the network to have multiple labels. In fact, we force all leaf labels to be conserved as the network is trimmed by cherry reductions by subsuming labels of a removed leaf onto its cherry sibling that remains. In this way, we keep a ``memory" of reductions and this compressed representation of networks allows to restore all possible alternative network (bijective) leaf labelings from it. 

Finally, we conclude the paper by discussing how the enumeration step could be optimized by considering the relationships between the reticulations of both input networks. We also briefly present a preliminary idea of how the proposed algorithm could be extended to higher level binary networks. 
Even though the proposed approach applies to orchard networks and not to general networks, the orchard network class actually contains network types that are of interest to the research community, such as the tree-child networks~\cite{bordewich2016determining} and tree-sibling time-consistent networks~\cite{erdHos2019class}. The tree-child networks in particular, in addition to having been studied extensively in the literature, are biologically relevant, since all ancestral species (internal vertices) have a path that can go to a leaf using only tree vertices. This reflects the idea that ancestral species have descendants that will perdure through mutation and speciation events, and that hybridization events are not as common as speciation events~\cite{kong2022classes}.

%The problem we address here is that of MACRS, first defined and shown to be NP-hard in the context of \emph{cherry distance}, a distance based on the \emph{cherry reduction} network trimming operation\cite{CDIST21,CDIST22}. In this previous work we find we can equivocate the two problems, and so here we refer to only one of them with the implied result that the strategy is applied towards solving the cherry distance problem. 

%\ots{add complexity, add FPT (retics)}

\section{Preliminaries}
        We first introduce the notions regarding networks, then proceed to defining cherry operations and our problem of interest.

\subsection{Networks}
\label{sss:nets}

A \emph{phylogenetic network} $\N$, or a \emph{network} for short, is an acyclic directed graph without vertices of in-degree and out-degree $1$, and whose vertices and edges are denoted $V(\N)$ and $E(\N)$, respectively.  We assume that all networks are
binary.  For $v \in V(\N)$, we use $v^-$ and $v^+$ to denote the in-degree and out-degree of $v$, respectively.
The set $V(\N)$ contains
\begin{itemize}
  \item the \emph{root} $\rho(\N)$, which is the unique node satisfying $\rho(\N)^-=0$ and $\rho(\N)^+=2$. In the case that $|V(\N)| = 2$, $\rho(\N)^+=1$;
  
  \item the \emph{leaves} $L(\N)$, which satisfy $l^-=1$ and $l^+=0$ for all $l \in L(\N)$;
  % labelled by some subset of $X(\N)$, and $\forall l\in L(\N)$, $l^-=1$, and $l^+=0$, note that label sets for all members of $L(\N)$ are disjoint;
  
  \item the \emph{internal} vertices $V(\N) \setminus (L(\N) \cup \{\rho(\N)\})$, which contains:
        \begin{itemize}
          \item the \emph{tree vertices} $T(\N)$, which satisfy $v^-=1$ and $v^+ = 2$ for all $v \in T(\N)$;
          
          \item the \emph{reticulation vertices} $R(\N)$, or simply \emph{reticulations}, which satisfy $v^- = 2$ and $v^+=1$ for all $v\in R(\N)$.
        \end{itemize}
\end{itemize}
We use $X$ to denote the set of all taxa.  For our purposes, the leaves of a network $\N$ are labeled by one \emph{or more} taxa.  For $l \in L(\N)$, we will use $X(l)$ to denote the set of taxa that label $l$.  We require that $X(l) \neq \emptyset$, and that for any distinct leaves $l_1, l_2 \in L(\N)$, $X(l_1) \cap X(l_2) = \emptyset$.  
%Note that there may be some labels of $X$ that are not in the label set of any leaf. 
%\comment{ML}{I suggest we just call them networks, and drop the `multi'}

The edges directed into a reticulation vertex are called \emph{reticulation
edges}, denoted $E_R(\N)$.  For $v \in V(\N)$, the out-neighbors of $v$ are called its \emph{children}.  If $v$ has a single in-neighbor, we denote it by $p(v)$ and call it the \emph{parent} of $v$ (if $v \in \{\rho(\N)\} \cup R(\N)$, then $p(v)$ is undefined).
Vertices $u$ and $v$ are \emph{siblings} if $p(u), p(v)$ are defined and $p(u) = p(v)$.
% The directed edges creates an ordering on the vertices where on edge $(u,v)$, $u$ is
% called the \emph{parent} of $v$ with the notation $u=p(v)$ (when $v$ is not reticulated), and $v$ is called
% the \emph{child} of $u$. 
When there is a directed path from vertex $v$ to vertex $u$, we call $v$ an \emph{ancestor} of $u$ and we call $u$ a \emph{descendant} of $v$. The descendants of $v$ are denoted $\below(v,\N)$ while its ancestors are denoted $\above(v,\N)$ (note that $v$ itself is in both sets). 
The union of the labels in $\below(v,\N) \cap L(\N)$ is denoted $X(v)$. 
We denote by $R(v)$ the set of reticulations in $\below(v,\N)$. 

Two networks $\N_1, \N_2$ are \emph{weakly isomorphic} if there exists a bijection $\sigma : V(\N_1) \rightarrow V(\N_2)$ such that $(u, v) \in E(\N_1)$ if and only if $(\sigma(u), \sigma(v)) \in E(\N_2)$, and such that for each $l \in L(\N_1)$, $X(l) \cap X(\sigma(l)) \neq \emptyset$. For this  we use the notation $\N \simeq \N'$. If, for each $l \in L(\N_1)$, $X(l) = X(\sigma(l))$, then we say $\N_1$ and $\N_2$ are \emph{strongly isomorphic} which we denote by $\N_1=\N_2$.

% Network edges are directed towards the leaves. Directed cycles and parallel edges are not
% permitted.  

%special networks

A network $\N$ may have only one edge whose endpoints are $\rho(\N)$ and a leaf. %\comment{ML}{In the def of network, the root has 2 children.  Should we change that?}
Then $\N$ is a \emph{single-leaf network} or \emph{singleton}. 
We say $\rho(\N)$ \emph{roots} $\N$. If, for a vertex $v$, and for all vertices $v' \in \below(v,\N)$, if  every path from $\rho(\N)$ to $v'$ goes through $v$, then we say $v$ roots the subnetwork below it.

While a network $\N$ is directed, there is an undirected version of $\N$ on the same vertex set and with an undirected edge $\{u,v\}$ present for every $(u,v)\in E(\N)$ which we call the \emph{underlying graph}. %$g(\N)$. 
It is on this underlying graph that we identify the set of \emph{biconnected components} of $\N$. Such a component is a maximal subgraph $B$ that cannot be disconnected by the removal of an edge therein. Note that every individual leaf and some tree vertices alone constitute a biconnected component, we refer to such  single vertex components as \emph{trivial}, and all others as \emph{non-trivial}. For a set of biconnected components $B_1...B_b$ on a network $\N$, a \emph{bridge} is an edge $(u,v)$ such that $u\in B_i$, $v\in B_j$ for any arbitrary $1\leq i \neq j \leq b$. 
%Intuitively, a bridge disconnects $g(\N)$ when removed. In a tree, every biconnected component is trivial (each vertex) and every edge is a bridge. In a network, biconnected components that are non-trivial contain a reticulation, they always have more than two vertices since parallel edges are not allowed.  \comment{ML}{This reminds me, if you contract each $B_i$ to a single vertex, you get a tree (edges are the bridges).  Sometimes, dynamic programming is useful on the tree of biconnected components.}

The \emph{level} of a network is the maximum number of reticulations across all biconnected components of a network. A level-$k$ network has no biconnected component with more than $k$ reticulations. A level-$1$ network has every biconnected component with either $0$ or $1$ reticulations. Note that this does not limit the number of reticulations over the whole network, just in each biconnected component. 

%%%%%%%%%%%%%%%%%%%%%%%%%%%%%%%%%%%%%%%%%%%%%%%%%%%%%%%%%%%%%%%%%
\subsection{Cherries and cherry reductions}

A \emph{cherry} is a pair of leaves that are siblings or that have a reticulation joining their parents.
%of sibling or, because an additional
%reticulation sits between them, nearly sibling leaves. 
More specifically, a pair $(x, y) \in L(\N) \times L(\N)$
is called a \emph{cherry} if either $p(x) = p(y)$, in which case $(x, y)$ is called a \emph{simple} cherry, 
or $p(x) \in R(\N)$ and $(p(y), p(x)) \in E(\N)$, in which case $(x, y)$ is called a \emph{reticulated} cherry.

%A \emph{cherry reduction} is a network operation
%performed on a cherry that exists in the network. 
Let $\N$ be a network and let $(x, y)$ be a pair of vertices.  Then \emph{applying the cherry reduction $(x, y)$} on $\N$ creates a new network as follows:
\begin{itemize}
    \item 
 If $(x, y)$ is a simple cherry of $\N$, then the \emph{$(x, y)$ reduction} consists of removing the leaf $x$ and
the edge $(p(x),x)$, suppressing the resulting node of in and out-degree $1$ if any, and re-assigning $X(y) = X(y)\cup X(x)$.  %\comment{ML}{need figure to illustrate this subsuming idea}
Note that the operation we introduce here differs from the cherry reduction operation described in previous work, where both the leaf $x$ and the set $X(x)$ are deleted. The purpose of our new definition is to preserve a reference to which label \emph{could} have been assigned to $y$. This is to say that the labels on a given leaf are interchangeable~\cite[Lemma~3]{CDIST22}.

%\ml{The purpose of merging the labels of $x$ and $y$ is to keep track of which label \emph{could} be assigned to $y$, see e.g. [REF to Thm saying any leaf could be chosen last].} 

    \item 
    If $(x,y)$ is a reticulated cherry of $\N$, then we remove the reticulation
edge $(p(y),p(x))$  and the resulting vertices of in and out-degree $1$ are suppressed.  In this case, we say that the reticulation edge $(p(y), p(x))$ is \emph{removed} by the cherry reduction $(x, y)$.

    \item 
    If $(x, y)$ is not a cherry of $\N$, then $\N$ is unchanged.
    
\end{itemize}

% If $p(x)^-=1$, it will not longer be a network vertex since $p(x)^+=1$ also, so it is identified with its parent. 

%Both $x$ and $y$ are identified with their parent. 
The resulting graph is a network, and always has a cherry unless it is a singleton (true of all orchard networks by definition \cite{erdHos2019class,janssen2021cherry}). 

%The set of cherries on a given network $\N$ are the \emph{available cherries}, $C(\N)$. 

Cherry reductions often occur in batches,  and a sequence $S$ of pairs of leaves is called a \emph{cherry sequence} (\emph{CS}).  The number of elements in $S$ is denoted $|S|$.
The cherry at position $i$ of a CS $S$ is referred to by $S_i$.  We use $\cs{\N}{S}$ to denote the network obtained from $\N$ by first applying cherry reduction $S_1$ on $\N$, then $S_2$ on the resulting network, and so on until $S_{|S|}$ is applied.  Note that we allow $S$ to contain pairs that do not modify the network (e.g. non-cherries).
The subsequence from (including) the first cherry to (excluding) the $i$th cherry in $S$ is $S_{(0:i)}$.  %\ml{[ML: todo: check if we need this notation?]}\kl{we do}
%or $S_{[1:i]}$. From the first to $i$th cherry, not including cherry $i$ is $S_{(0:i)}$. 
% When a network $\N$ undergoes the ordered sequence of reductions in a CS $S$, we say $S$ is \emph{applied} to $\N$ and use the notation $\cs{\N}{S}$. 
When a CS $S$ reduces a network $\N$ to a singleton, then we say $S$ is \emph{complete} for $\N$. 
%A network is an \emph{orchard} if there exists a CS that is complete for $\N$\cite{erdHos2019class}. 
We assume networks are orchard networks hereafter.
%$|S|$ denotes the number of cherries in a CS $S$. \ml{[ML: $|S|$ already mentioned above, watch out for these]}.  
%For CS $S$ on a network $\N$ such that there is no $i$ where $\cs{\N}{S_{(0:i)}} = \cs{\N}{S_i}$, we call $S$ \emph{minimal}. \ml{[ML: do we need minimal? ]}

See Figure~\ref{fig:reductions} for an illustration of the two cherry reduction operations, and the concepts of isomorphism. 
\def\arrow{
  \coordinate (l) at (0,1.75);
  \coordinate (r) at (1,1.75);
  \draw[->] (l) -- (r);
  \coordinate (ghost) at (0,0);
  \node at (ghost) {};
}
\def\basetree{
    \coordinate (root) at (0,0);
    \foreach \v/\i in {a/-2,r1/0,b/2} {
        \coordinate (\v) at (\i,-2);
    }
    \node[leaf,label={[xshift=1mm,yshift=-8mm]\small{$a=\{a\}$}}] at (a){};
    \node[leaf,label={[xshift=1mm,yshift=-8mm]\small{$b=\{b\}$}}] at (b){};
    \path (r1) +(90:0.5) coordinate (r1n);
    \path (r1) +(-90:0.5) coordinate (root2);
    \node[ret,label={[xshift=4mm,yshift=-4mm]\small{$r_1$}}] at (r1n){};
    \node[tree] at (root2){};
    \draw (r1n) -- (root2);
    \foreach \v/\i in {c/-2,r2/0,f/2} {
        \coordinate (\v) at (\i,-4);
    }
    \node[leaf,label={[xshift=1mm,yshift=-8mm]\small{$c=\{c\}$}}] at (c){};
    \node[leaf,label={[xshift=1mm,yshift=-8mm]\small{$f=\{f\}$}}] at (f){};
    \draw (a) -- (root) -- (b);
    \draw (c) -- (root2) -- (f);
    \path (r2) +(90:0.5) coordinate (r2n);
    \coordinate (u) at (barycentric cs:root=0.6,a=0.5);
    \coordinate (v) at (barycentric cs:root=0.6,b=0.5);
    \coordinate (w) at (barycentric cs:root2=0.6,c=0.5);
    \coordinate (x) at (barycentric cs:root2=0.6,f=0.5);
    \node[tree,label={[xshift=-2mm,yshift=-1mm]$u$}] at (u){};
    \node[tree,label={[xshift=2mm,yshift=-1mm]$v$}] at (v){};
    \node[tree,label={[xshift=-3mm,yshift=-3mm]\small{$w$}}] at (w){};
    \draw (u) -- (r1n) -- (v);
}
\def\oneandtwo{
    \draw (w) -- (r2n);
    \node[tree,label={[xshift=3mm,yshift=-3mm]\small{$x$}}] at (x){};
    \draw (x) -- (r2n);
    \node[ret,label={[xshift=4mm,yshift=-5mm]\small{$r_2$}}] at (r2n){};
}
\def\basetreetwo{
    \coordinate (root) at (0,0);
    \coordinate (a) at (-1,-1.5);
    \coordinate (b) at (1,-1.5);
    \node[leaf,label={[xshift=1mm,yshift=-8mm]\small{$a=\{a\}$}}] at (a){};
    \node[leaf,label={[xshift=1mm,yshift=-8mm]\small{$b=\{b\}$}}] at (b){};
    \coordinate (u) at (barycentric cs:root=0.6,a=0.5);
    \coordinate (v) at (barycentric cs:root=0.6,b=0.5);
    \node[tree] at (u){};
    \coordinate (c) at (0,-2);
    \draw (a) -- (root) -- (b);
    \draw (u) -- (c);
}
\begin{figure}
    \centering
    \resizebox{0.80\linewidth}{!}{
    \begin{tikzpicture}
        \basetree
        \node[root,label={[xshift=-1.8cm,yshift=0mm]\Large{$\N_1$}}] at (root){};
        \oneandtwo
        \node[tree] at (r2){};
        \draw (r2n) -- (r2);
        \coordinate (d) at (-1,-4.5);
        \coordinate (e) at (1,-4.5);
        \node[leaf,label={[xshift=1mm,yshift=-8mm]\small{$d=\{d\}$}}] at (d){};
        \node[leaf,label={[xshift=1mm,yshift=-8mm]\small{$e=\{e\}$}}] at (e){};
        \draw (d) -- (r2) -- (e);
    \end{tikzpicture}
    \hspace{1em}%--------------------------------------
    \vspace{1cm}%--------------------------------------
    %\begin{tikzpicture}
        %\arrow
        %\coordinate (text) at (0.5,2.25);
        %\node[anchor=base,align=center] at (text)%{$(d,e)$};
%\end{tikzpicture}
    \hspace{1em}%--------------------------------------
    \begin{tikzpicture}
        \basetree
        \node[root,label={[xshift=-1.8cm,yshift=0mm]\Large{$\N_2$}}] at (root){};
        \oneandtwo
        \node[leaf,label={[xshift=1mm,yshift=-8mm]\small{$e=\{d,e\}$}}] at (r2){};
        \draw (r2n)--(r2);
    \end{tikzpicture}
    }
    \resizebox{0.90\linewidth}{!}{
    %\vspace{4em}
    \hspace{1em}%--------------------------------------
    %\begin{tikzpicture}
        %\arrow
        %\coordinate (text) at (0.5,2.25);
        %\node[anchor=base,align=center] at (text)%{$(e,f)$};
    %\end{tikzpicture}
    \hspace{1em}%--------------------------------------
    \begin{tikzpicture}
        \basetree
        \node[root,label={[xshift=-1.8cm,yshift=0mm]\Large{$\N_3$}}] at (root){};
        \node[leaf,label={[xshift=1mm,yshift=-8mm]\small{$e=\{d,e\}$}}] at (r2){};
        \draw (w) -- (r2);
    \end{tikzpicture}
    \hspace{1em}%--------------------------------------
    \begin{tikzpicture}
        \basetreetwo
        \node[root,label={[xshift=-1cm,yshift=0mm]\Large{$\N_4$}}] at (root){};
        \node[leaf,label={[xshift=1mm,yshift=-8mm]\small{$e=\{c,d,e,f\}$}}] at (c){};
    \end{tikzpicture}
    \hspace{1em}%--------------------------------------
    \begin{tikzpicture}
        \basetreetwo
        \node[root,label={[xshift=-1cm,yshift=0mm]\Large{$\N_5$}}] at (root){};
        \node[leaf,label={[xshift=1mm,yshift=-8mm]\small{$c=\{c\}$}}] at (c){};
    \end{tikzpicture}
    }
    \caption{In this figure, leaves are represented by open circles, tree vertices as filled circles, reticulations as filled squares, and the root of the network as a filled, inverted triangle. Network $\N_1$ is a level-$1$ network with $|R(\N)| = 2 $. $\N_1$ is a reticulation-trimmed subnetwork of $\N_1$ with respect to $F=\emptyset$. Network $\N_2 = \cs{\N_1}{(d,e)}$, where $(d,e)$ is a simple cherry/reduction. Network $\N_3 = \cs{\N_2}{(e,f)}$ where $(e,f)$ is a reticulated cherry/reduction. $\N_3$ is reticulation-trimmed subnetwork of $\N_1$ and of $\N_2$ with respect to $F =\{ (x,r_2) \}$. Network $\N_4 = \cs{\N_3}{(c,e)\cdot(f,e)\cdot(e,b)}$ and is a reticulation-trimmed subnetwork of $\N_1$ and of $\N_2$ with respect to $F = \{ x,r_2), (v,r_1) \}$ or to $F = \{ (w,r_2), (v,r_1) \}$. Network $\N_5 \simeq \N_4$, in fact, there are CSs that may head lead to leaf $e$ being any of leaves $c$, $d$, $e$, or $f$. Each of these networks would have the same label set on that leaf, and all are weakly isomorphic with $\N_5$.
    }
% For each series of cherry reductions that lead to a structure like (4), there may be a different leaf in place of $e$ (any of c,d,e,f), the label set of any such leaf will remain as pictured, all the reductions are strongly isomorphic. 
    \label{fig:reductions}
\end{figure}
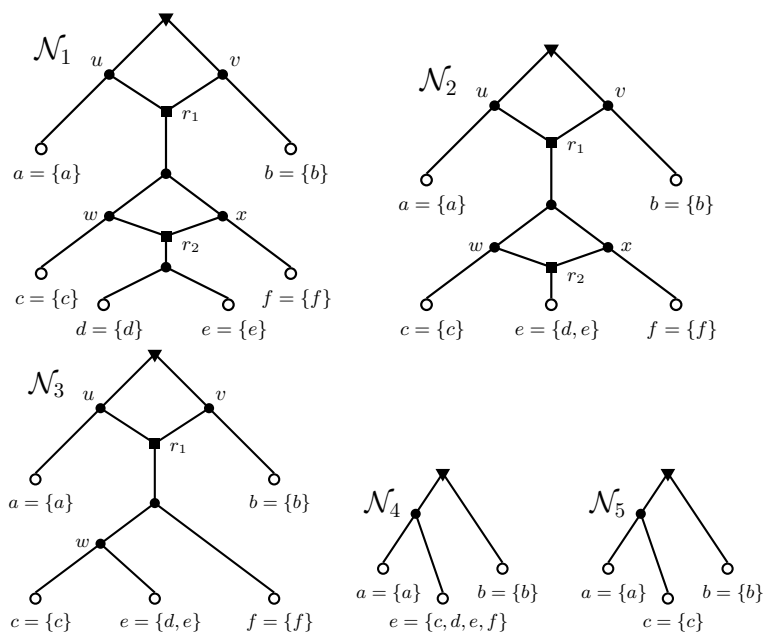

Cherries on a network can be reduced in any order. %\cite[Theorem~9]{janssen2020linear} 
We restate a theorem of~\cite{janssen2020linear} that we adapt to our formalism\footnote{Note that the authors prove the statement under the assumption that $S$ is complete, and that leaves are single-labeled.  However the proof is easy to adapt to our context.}.  

\begin{restatable}{theorem}{anyorder}
    \label{thm:any-order}
    Let $\N$ be a network, let $(x,y)$ be a cherry of $\N$, and let $S$ be a CS that contains $(x, y)$.  Then there exists a CS $S'$ such that $\cs{\N}{S} = \cs{\N}{S'}$, and whose first element is $(x,y)$.
    %\ml{[ML: did we define what a minimal CS of $\N$ is?]}KL put this inder a definition of minimal
\end{restatable}
%original wording: Let N be a tree-child network and let (x, y) be a reducible pair of N . Then there exists a minimal TCS of N whose first element is (x, y

%\comment{ML}{This section is a bit long, especially for a conf version.  I'm sure it's possible to make it more succinct using the notation.}

%For a given leaf $l\in \cs{\N}{S}$ for some network $\N$ and CS $S$, we define the \emph{subsume} set of $l$ on $\cs{\N}{S}$ as $subsume(l,S)$. This is a set of leaf labels that $l$ has been in a cherry reduction with. More precisely, $subsume(l,S_0)=\{l \}$, $subsume(l,S_{(0:i]})=subsume(l,S_{(0:i)})\cup subsume(m,S_{(0:i)})$ if $S_i=(l,m)$, $subsume(l,S_{(0:i]})=subsume(l,S_{(0:i)})$ otherwise.

\subsection{Maximum agreement cherry-reduced subnetworks}

For networks $\N$ and $\N'$, when there exists a CS $S$ such that $\cs{\N}{S} \simeq \N'$, we say that $\N'$ is a \emph{cherry-reduced subnetwork} (\emph{CRS}) of $\N$, denoted by $\N' \crof \N$. We can now define the main problem of focus. 
%For networks $\N_1$, $\N_2$, and $\N^*$, such that $\N^*$ is a CRS of both $\N_1$ and $\N_2$, we say that $\N^*$ is an \emph{agreement} cherry-reduced subnetwork, an ACRS, of $\N_1$ and $\N_2$. The next problem definition extends these definitions even further. 

\vspace{3mm}
\noindent
The \emph{Maximum Agreement Cherry-Reduced Subnetwork} (\macrs) problem. \\
\noindent
\textbf{Input:} Two orchard networks $\N_1$ and $\N_2$ \\
\noindent
\textbf{Find:} A network $\N^*$ with the maximum number of vertices that satisfies $\N^* \crof \N_1$ and $\N^* \crof \N_2$ 
%\kl{and such that $X(\N_1)\cap \N_2 \neq \emptyset$}.
\vspace{3mm}

A solution $\N^*$ to the above problem will be called an \macrs~of $\N_1$ and $\N_2$.

% This implies the existence of CSs $S_1$ and $S_2$ such that $\cs{\N_1}{S_1} \simeq \cs{\N_2}{S_2}$, note how we only require weak isomorphism for this problem.
% -----------------
%\section{Problem Definitions}
 %       \input{prob-def}
% -----------------     
\section{An MACRS algorithm on level-1 networks}
    %tex master:  main
%reduced set, then macrs-simple, then main alg and proofs

%
%
%The conjecture Algorithm in thoery

We show that the MACRS problem can be solved in time $O(3^r n^3)$ for $n = \max(|V(\N_1)|, |V(\N_2)|)$, and $r = |R(\N_1)| + R(\N_2)|$ on level-$1$ networks.
% All networks referred to hereafter are assumed to be level-$1$ unless otherwise stated. 
We employ a two-step strategy.  We first enumerate a number of inputs that have been specially reduced to a selected set of remaining reticulations. Second, these inputs are provided to a cubic time dynamic programming algorithm on an easier version of MACRS that uses only simple reductions. Because of the number of special inputs is limited by $3^r$, we get an FPT 
%\ml{[ML: make sure that acronym is defined somewhere]} 
algorithm. 
\macrs~is thus split into two subproblems.  We first introduce them and show how they can be used to solve \macrs.  The later sections then focus on each problem separately.

%
%\subsection{subroutine 1}
%define the first subproblem: reduced set

Let $\N$ be a network and let $F \subseteq E_R(\N)$ be a subset of reticulation edges.  We wish to generate all the maximal cherry-reduced subnetworks of $\N$ under the restriction that the reticulation edges removed by cherry operations coincide with $F$.
% that result from choosing whether to keep or remove each reticulation in the network. Furthermore, when a reticulation is removed, the reduction can proceed on one of the two corresponding reticulation edges. We endeavor to keep reductions minimal so that all smaller CRS are kept intact to be discovered by the MACRS-SIMPLE subroutine. Thus we land on the following definition of the special subnetwork we wish to enumerate. 
%
Thus, we say that a network $\N'$ is a \emph{reticulation-trimmed subnetwork} of $\N$ \emph{with respect to $F$} if there
exists
    a CS $S$ such that $\cs{\N}{S} = \N'$, 
    and such that $(u, v) \in F$ if and only if 
    $S$ contains a reticulated cherry reduction that removes $(u, v)$, and $S$ is of minimum length i.e. we require that there is no other CS $S'$ with $|S'| < |S|$ that satisfies the same properties.

%\end{itemize}

Furthermore, we say that $\N'$ is a \emph{reticulation-trimmed subnetwork} of $\N$ if there exists a set $F \subseteq E_R(\N)$ such that $\N'$ is a reticulation-trimmed subnetwork of $\N$ with respect to $F$.

\vspace{3mm}
\noindent
The \rettrimmed~problem: \\
\noindent
\textbf{Input:} An orchard network $\N$.\\
\noindent
\textbf{Find:} the set of all reticulation-trimmed subnetworks of $\N$.
\vspace{3mm}

%\comment{KL}{we can give the exact size of this set, background required}
%\comment{ML}{That seems difficult.  Could be stated as an open problem.}
%\comment{KL}{but i already solved with the dependency trees? last ald in old/old-gen-input}
%\comment{ML}{yes, there's an algorithm for it, but I had a close formula in mind in my comment.  I agree we can allude to the idea of computing it}
%\comment{KL}{closed form meaning in terms of the network itself? rather than in terms of dependency tree/branches?}
%The total size of this set is bounded by $3^{|R(\N)|}$\comment{KL}{we can count this exactly by defining the dependancy trees} ($2^{|E_R(\N)|}$ a naive but incorrect upper bound), but it is worth noting that not every subset of $E_R(\N)$ admits a reticulation-trimmed subnetwork;

Note that the size of the set of reticulation-trimmed subnetworks depends heavily on the network structure.  For instance, it is possible to show that it is linear when all reticulations are arranged in a path, and exponential when all reticulations are independent (none is an ancestor of the other).
It is possible to calculate the size of this set exactly by algorithmic means though an abstraction of the network structure.  However, we reserve the analysis of the impact of this parameter on our algorithm for future work.
%a closed form formula for this counting problem is an open problem.}

%
%\subsection{subroutine 2}
%define the second subproblem: macrs-simple

Once the set of edges to remove by reticulation have been guessed, it remains to infer the set of non-reticulated cherry operations.
A \emph{simple CS} is a CS that contains only simple cherries.  In this way, $R(\N)=R(\cs{\N}{S})$ for any simple CS $S$. For networks $\N$ and $\N'$, when there exists a simple CS $S$ such that $\cs{\N}{S} \simeq \N'$ we say that $\N'$ is a CRS-SIMPLE of $\N$.
Note that owing to our definition of weak isomorphism, $\cs{\N}{S} \simeq \N'$ does not mean that $S$ transforms $\N$ into $\N'$.  A better intuition would rather be that after applying $S$ on $\N$, we could choose one label in the label set of each leaf of $\cs{N}{S}$ and of $\N'$, such that the resulting networks would be isomorphic in the traditional sense.

\vspace{3mm}
\noindent
The \emph{Simple Maximum Agreement Cherry-Reduced Subnetwork} (\macrssimple) problem. \\
\noindent
\textbf{Input:} Two orchard networks $\N_1$ and $\N_2$. \\
\noindent
\textbf{Find:} a network $\N^*$ with a maximum number of vertices such that $\N^*$ is a CRS-SIMPLE of $\N_1$ and a CRS-SIMPLE of $\N_2$.

\vspace{3mm}

A solution $\N^*$ to the above problem will be called a \macrssimple~of $\N_1$ and $\N_2$.

For the standard MACRS problem on networks $\N_1$ and $\N_2$, there is always a solution as long as $X(\N_1)\cap X(\N_2)\neq \emptyset$ %\comment{ML}{did we?}
, however since reticulations cannot be removed by simple CS, the MACRS-SIMPLE problem may not have a solution (for instance when the two networks have different number of reticulation vertices). 
We can now describe our main algorithm, where we assume that the \macrssimple~routine correctly returns an optimal solution to the above problem.
%\ml{[ML: I suggest the following algo, which doesn't require more definitions, and doesn't require defining routines that are in later sections.]}

\begin{singlespace}
\begin{algorithm}[H]
    \caption{MACRS Finder}\label{alg:macrs}
    \hspace*{\algorithmicindent} \textbf{Input} Two networks $\N_1$ and $\N_2$ \\
    \hspace*{\algorithmicindent} \textbf{Output} A MACRS of $\N_1$ and $\N_2$
    \begin{algorithmic}[1]
        \State $\tilde{\N} \gets $~empty network
        \ForEach{reticulation-trimmed subnetwork $\N'_1$ of $\N_1$}
            \ForEach{reticulation-trimmed subnetwork $\N'_2$ of $\N_2$}
                \State Let $\N'$ be a \macrssimple~of $\N'_1$ and $\N'_2$
                %below is a fake if then statement on one line cuz idk how to do
                \State \textbf{if} $\N'$ exists and $|V(\N')| > |V(\tilde{\N})|$ \textbf{then} $\tilde{\N} \gets \N'$
            \EndFor
        \EndFor
        \State return $\tilde{\N}$
    \end{algorithmic}
\end{algorithm}
\end{singlespace}

%\ml{[ML: say somewhere that if $\N_1$ and $\N_2$ have different number of reticulations, we can skip them.]}
An optimization technique is evident here: as we mentioned, there is only a solution to \macrssimple~($\N_1$, $\N_2$) when $|R(\N_1)| = |R(\N_2)|$ since only simple reductions will be performed. Thus, we need only test such pairs. This optimization is not currently formalized into the algorithm and complexity analysis presented here, but rather will make for future work. 

In the remainder of this section, we focus on proving that this algorithm works correctly.  We will deal with the complexity of the algorithm once we have dealt with the \rettrimmed~and \macrssimple~subproblems.
We begin by showing that one can always obtain a subnetwork by first going through a reticulated-trimmed subnetwork, and then using only simple cherry reductions. 

\begin{restatable}{lemma}{subthroughtrim}\label{lem:sub-through-trim}
Let $\N$ be a network.  Then 
for any $\N' \crof \N$, there exists a reticulation-trimmed subnetwork $\N''$ of $\N$ and a simple CS $S$ such that $\cs{\N''}{S} = \N'$.
\end{restatable}

For proof of Lemma~\ref{lem:sub-through-trim}, see Appendix. 

%
%\subsection{main algorithm}
%formal outline of conjecture algorithm 

% \begin{singlespace}
% \begin{algorithm}[H]
%     \caption{MACRS Finder}\label{alg:macrs}
%     \hspace*{\algorithmicindent} \textbf{Input} Two multi-networks $\N_1$ and $\N_2$ \\
%     \hspace*{\algorithmicindent} \textbf{Output} MACRS($\N_1$,$\N_2$)
%     \begin{algorithmic}[1]
%         \State $\textbf{A} \leftarrow$ REDUCED-SET($\N_1$), $\textbf{B} \leftarrow$ REDUCED-SET($\N_2$) 
%         \State $\N \leftarrow \text{max}(\text{MACRS-SIMPLE}(A,B)) \forall (A,B)\in \textbf{A} \times \textbf{B}$
%         \State return $\N$
%     \end{algorithmic}
% \end{algorithm}
% \end{singlespace}
% \comment{KL}{referencing the problems not the algorithms?}
%
%
%basic proof of conjectured alg 

\begin{restatable}{theorem}{macrscorrect}
\label{thm:macrs}
    Algorithm~\ref{alg:macrs} correctly finds a \macrs~of $\N_1$ and $\N_2$.
    %\ml{[ML: I removed the complexity since, at this stage, we are not in a position to analyze it.  We can leave the final complexity analysis for the very end in my opinion.  The current statement could be turned into a lemma (or thm either way).]}
    %returns $\N$ for $\N = MACRS(\N_1,\N_2)$ in $O(3^{|R(\N_1)|+|R(\N_2)|})$.
\end{restatable}
\begin{proof}
    %\ml{[ML: I restructured this proof because it doesn't need anything sophisticated: Lemma 2 does all the job.]}
    Let $\N^*$ be a MACRS of $\N_1$ and $\N_2$.  
    Let $\tilde{\N}$ be the network returned by Algorithm~\ref{alg:macrs}.  We first claim that if $\tilde{\N}$ is non-empty, it does satisfy $\tilde{\N} \crof \N_1, \N_2$.
    To see this, note that every pair $\N'_1, \N'_2$ of networks enumerated by Algorithm~\ref{alg:macrs} satisfy $\N'_1 \crof \N_1$ and $\N'_2 \crof \N_2$, by the definition of reticulation-trimmed subnetworks.
    Moreover, if a \macrssimple~$\N'$ of $\N'_1, \N'_2$ exists, then by transitivity, $\N' \crof \N'_1 \crof \N_1$ and $\N' \crof \N'_2 \crof \N_2$.  Since $\tilde{\N}$ is one of those $\N'$, this proves our claim.
    
    Let us now focus on the optimality of $\tilde{\N}$.
    First note that $|V(\N^*)| \geq |V(\tilde{\N})|$: if $\Tilde{\N}$ is an empty network, this is obvious, and otherwise, by our above claim, $\tilde{\N}$ is a cherry-reduced subnetwork of $\N_1$ and $\N_2$ and can thus not be larger than $\N^*$.

    Let us now show that $|V(\N^*)| \leq |V(\tilde{\N})|$.
    By Lemma~\ref{lem:sub-through-trim}, there exists a reticulation-trimmed subnetwork $\N'_1$ of $\N_1$ (resp. $\N'_2$ of $\N_2$) such that $\N^*$ can be obtained from $\N'_1$ (resp. $\N'_2$) using only simple CSs.  Thus, $\N^*$ is a CRS-SIMPLE of $\N'_1$ and $\N'_2$.  Algorithm~\ref{alg:macrs} will eventually enumerate $\N'_1$ and $\N'_2$ and find a \macrssimple~$\N'$ of them, which is of maximum size and thus has at least as many vertices as $\N^*$.  Since the returned $\tilde{\N}$ is the $\N'$ of maximum size found by the algorithm, it follows that $|V(\N^*)| \leq |V(\tilde{\N})|$.

\end{proof}

% -----------------
\section{Subroutines}
    %tex master: main

    \subsection{Enumerating the set of reticulation-trimmed subnetworks}
        \label{ss:reduced-set}
        %tex master: lvl 1 < main
\label{ss:r-t-subnet-enum}
We now show how to enumerate the set of all 
reticulation-trimmed subnetworks of a network $\N$ in time $O(3^{|R(\N)|}|V(\N)|)$.
The reticulation-trimmed subnetworks are characterized by having no more reductions than what sufficiently removes the desired reticulation edges. Luckily, we will see that at most one such network can exist; we must only remove the complete subnetwork under both endpoints of the reduced reticulation edge. This is guaranteed possible by cherry reductions, assuming all reticulations below these endpoints have also been specified for removal. 
Algorithm~\ref{alg:reduced-set} shows how to enumerate the relevant edges, and uses Algorithm~\ref{alg:make-rt-subnet} as a subroutine, which finds the reticulation-trimmed subnetwork with respect to a given edge set.
We show that the reticulation-trimmed subnetwork of $\N$ with respect to $F\subseteq R(\N)$ is uniquely defined in Lemma~\ref{lem:unique-rt-sub}.  We say that a set of edges $F$ is \emph{disjoint} if, for any two distinct edges $(u, v), (x, y) \in F$, $\{u, v\} \cap \{x, y\} = \emptyset$.

%[ A naive solution would seemingly produce $3^{R(\N)}$ networks, however many of those networks would not be valid (or would be duplicates, depending on how we generate them). To elucidate this point, its enough to say that we cannot decide to remove reticulations which are above reticulations which are decided to be kept intact, due to the bottom-up nature of cherry reductions. \comment{KL}{better context when discussing optimizations} ]

%\ml{[ML: algo 2 does not do a lot.  We could think of merging algo 2 and 3 (just a suggestion)]}

\begin{singlespace}
\begin{algorithm}[H]
    \caption{REDUCED-SET-FINDER}\label{alg:reduced-set}
    \hspace*{\algorithmicindent} \textbf{Input} A network $\N$ \\
    \hspace*{\algorithmicindent} \textbf{Output} The set of all reticulation-trimmed subnetworks of $\N$
    \begin{algorithmic}[1]
        \For{\textbf{each} $F \in \{\mathcal{P}(E_R(\N)) : (a,b), (c,b) \notin F$ for any $a,b,c$\}}
        %\ml{ML: not sure, shouldn't it be $(a, b), (c, b) \notin F$? (or both)KL:yes, updated}
            \State $\textbf{N} \leftarrow \textbf{N} \cup \Call{RT-SUBNET-MAKER}{\N,F}$
        \EndFor 
        \State return \textbf{N}
    \end{algorithmic}
\end{algorithm}
\end{singlespace}

%\ml{[Ok so I've never seen this way of formulating a ``Let''.  I'm used to ``Let $\N$ be a network'' instead of ``Let there be network $\N$''.  I don't really mind if you've seen it elsewhere, but what matters most is homogeneity, at least in the statements.  If time allows it, we need to make sure that we used the same formulation throughout.]}KL will adjust if time permits

\begin{restatable}{lemma}{disjointF} 
    \label{lem:disjoint-F}
    Let $\N$ be a network, $F \subseteq E_R(\N)$ be a set, and $\N'$ be a network that is a reticulation-trimmed subnetwork of $\N$ with respect to $F$. Then $F$ is disjoint.
\end{restatable}

For proof of Lemma~\ref{lem:disjoint-F}, see Appendix. 
Next, for $F \subseteq E_R(\N)$, a topological sort of $F$ is an ordering of its element such that for distinct edges $e_1, e_2 \in F$, if there is a path from a vertex of $e_1$ to a vertex of $e_2$ in $\N$, then $e_1$ comes later than $e_2$ in this ordering.

\begin{restatable}{lemma}{topsort}
    \label{lem:topsort}
    Let $\N$ be a network and $F \subseteq E_R(\N)$ be a set such that there exists a reticulation-trimmed subnetwork of $\N$ with respect to $F$.  Then there exists a topological sort of $F$.  
    %There is a topological sort on $F$ that produces a partial order.
\end{restatable}

For proof of Lemma~\ref{lem:topsort}, see Appendix. 
The next lemma is crucial, as it shows that reticulation-trimmed subnetworks with respect to a given $F$ are either unique, or do not exist.  This allows us to enumerate in reasonable time.

\begin{restatable}{lemma}{onertsubnet}
\label{lem:unique-rt-sub}
Let $\N$ be a network and let $F \subseteq E_R(\N)$.  Then there does not exist two non-strongly isomorphic reticulation-trimmed subnetworks of $\N$ with respect to $F$.
%Then there is either $0$ or $1$ reticulation-trimmed subnetworks of $\N$ with respect to $F$.

%\ml{Maybe restate to: there does not exist two non-strongly isomorphic reticulation-trimmed subnetworks of $\N$ wrt $F$.}
\end{restatable}
For proof of Lemma~\ref{lem:unique-rt-sub}, see Appendix. 
For an example, given a network $\N$, of an $F \subseteq E_R(\N)$ that does not admit a reticulation-trimmed subnetwork of $\N$, consider $\N$ with 2 reticulations, $r_1$, $r_2$ such that $r_1 \in \above(r_2)$. Choosing $F= \{ (p_1,r_1) \}$, for $p_1$ chosen arbitrarily between $r_1$'s parents, will not admit a reticulation-trimmed subnetwork since reticulation $r_2$ must have leaves below its endpoints to be in a cherry, but this choice of $F$ has no corresponding reticulated reductions of $r_1$ making it impossible to construct a CS $S$ that reduces only $r_2$.

We next describe Algorithm~\ref{alg:make-rt-subnet}, which produces the reticulation-trimmed networks with respect to some given $F$, see Figure~\ref{fig:r-t-subnet} for an illustration.

%----------------------------------Figure
\def\basetree{
    \coordinate (root) at (0,0);
    \node[tree,label={[xshift=6mm,yshift=-4mm]\large{$p(u)$}}] at (root){};
    \path (root) +(90:0.7) coordinate (above);
    \draw[decorate,decoration={snake,amplitude=0.7mm}] (root) -- (above);
    \foreach \v/\i in {a/-3,v/-1,v'/1,n/3}{
        \coordinate (\v) at (\i,-3);
    }
    \node[anchor=north,regular polygon, regular polygon sides=3,draw,inner sep=3mm] at (a){};
    \draw (a) -- (root);
    \path (a) (barycentric cs:a=0.9,root=0.3) coordinate (u);
    \node[tree,label={[xshift=-3mm,yshift=-3mm]\large{$u$}}] at (u){};
    \node[ret,label={[xshift=3mm,yshift=-5mm]\large{$v$}}] at (v){};
    \node[leaf,label={[xshift=3mm,yshift=-5mm]\large{$v'$}}] at (v'){};
    \node[decorate,decoration={snake,amplitude=0.3mm},draw,circle,inner sep=5mm,anchor=north,yshift=-0.5mm] at (n){};
    \path (v) +(-90:1) coordinate (vc);
    \draw (v)--(vc);
    \node[anchor=north,regular polygon, regular polygon sides=3,draw,inner sep=3mm] at (vc){};
    \coordinate (u') at (-1,-2);
    \node[leaf,label={[xshift=-3mm,yshift=-3mm]\large{$u'$}}] at (u'){};
    \draw (root) -- (n);
    \coordinate (pv) at (barycentric cs:root=0.8,n=0.3);
    \node[tree,label={[xshift=5mm,yshift=-2mm]\large{$p(v)$}}] at (pv){};
    \draw (pv) -- (v);
}
\begin{figure}[h]
    \centering
    \resizebox{1\linewidth}{!}{
    \begin{tikzpicture}
        \basetree
        \node[label={[xshift=-1.5cm,yshift=-7mm]\Large{(1)}}] at (root){};
        \draw[dashed] (u) -- (v);
    \end{tikzpicture}\
    \hspace{1em}%--------------------------------------
    \begin{tikzpicture}
        \basetree
        \node[label={[xshift=-1.5cm,yshift=-7mm]\Large{(2)}}] at (root){};
        \draw (u') -- (root);
        \draw  (pv) -- (v');

    \end{tikzpicture}
    %\hspace{1em}%--------------------------------------
    }
    \resizebox{0.35\linewidth}{!}{
    \begin{tikzpicture}
        \coordinate (root) at (0,0);
        \node[label={[xshift=-1.5cm,yshift=-7mm]\Large{(3)}}] at (root){};
       \path (root) +(90:0.7) coordinate (above);
        \draw[decorate,decoration={snake,amplitude=0.7mm}] (root) -- (above);
        \foreach \v/\i in {u/-2,v/0,n/2} {
            \coordinate (\v) at (\i,-2);
        }
        \node[leaf,label={[yshift=-7mm]\large{$u'$}}] at (u){}; 
        \node[leaf,label={[yshift=-7mm]\large{$v'$}}] at (v){}; 
        \draw (u) -- (root) -- (n);
        \coordinate (pv) at (barycentric cs:root=0.6,n=0.5);
        \draw (pv) -- (v);
        \node[decorate,decoration={snake,amplitude=0.3mm},draw,circle,inner sep=5mm,anchor=north,yshift=-0.5mm] at (n){};
    \end{tikzpicture}
    }
    %\vspace{-1em}
    \caption{In this figure, leaves are represented by open circles, tree vertices as filled circles, reticulations as filled squares. A subnetwork without reticulations is represented by a large open triangle, a subnetwork that may be reticulated is represented by a large open blob. This Figure shows an example of the operation of Algorithm~\ref{alg:make-rt-subnet}, note how $R(u)\cup R(v) \setminus\{v\} = \emptyset$ in this example. Subnetwork under label (1) is an example network at line 7, the dotted line represents the removed reticulation edge $(u,v)$ by line 5 and both leaves $u'$ and $v'$ have been constructed (leaf labels are not shown). The network under label (2) shows the state of network (1) at line 8 when edges $(p(u),u')$ and $(p(v),v')$ have been added.The network under label (3) shows the state of the network under (1) at line 9 when vertices in $\below(u,\N')\cup \below(v,\N')$ are removed. 
    \label{fig:r-t-subnet}
    }
\end{figure}
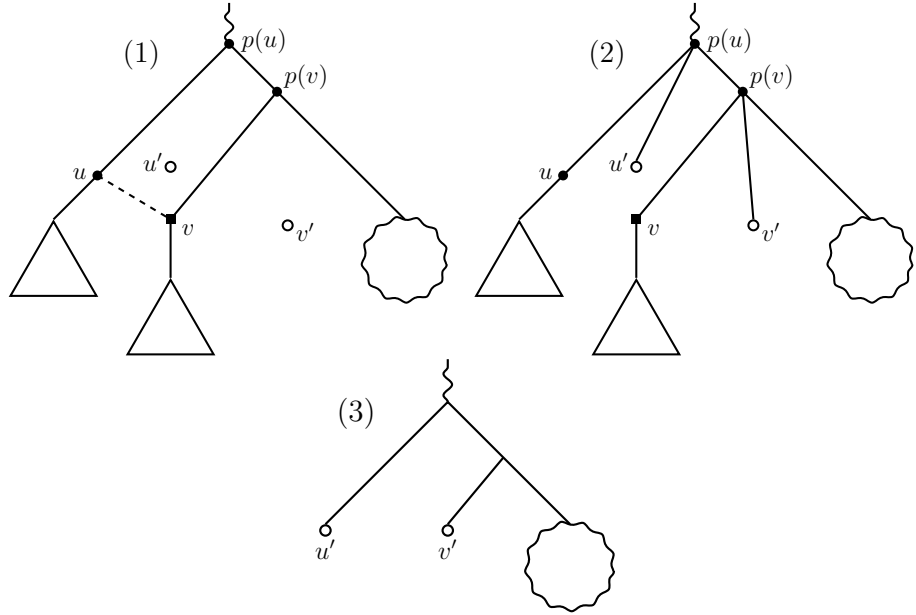

\begin{singlespace}
\begin{algorithm}[h]
    \caption{RT-SUBNET-MAKER}\label{alg:make-rt-subnet}
    \hspace*{\algorithmicindent} \textbf{Input} A network $\N$ and a disjoint set $F \subseteq E_R(\N)$\\
    \hspace*{\algorithmicindent} \textbf{Output} the reticulation-trimmed subnetwork of $\N$ with respect to $F$, or NULL if it does not exist 
    \begin{algorithmic}[1]
        \State $\N' \leftarrow \N$
        \State Find a topological sort $F'$ or $F$
        \For{\textbf{each} $(u,v) \in F'$ in order}
            \If{$R(u) \cup R(v) \setminus \{ v \} = \emptyset$}
            %\ml{ML: not valid notation.  $R(a) \in \N'$ is a boolean expression, so we can't use union $\cup$ with it.  Just write $R(a) \cup R(b)$.  Also, since $b$ is a reticulation, it belongs to $R(b)$ and so $R(a) \cup R(b)$  won't be empty.}  
                \State delete edge $(u,v)$
                \State construct leaf $u'$ such that $X(u') = X(u)$
                \State construct leaf $v'$ such that $X(v') = X(v)$
                \State add edges $(p(u), u')$ and $(p(v),v')$ to $\N'$
                \State remove all vertices in $\below(u,\N') \cup \below(v,\N')$
                %\State \ml{ML: we need to remove every vertex in $reach(a)$ and $reach(b)$, otherwise this is just a disconnected network}
            \Else
                \State return NULL
            \EndIf
        \EndFor
        \State return $\N'$
    \end{algorithmic}
\end{algorithm}
\end{singlespace}

%It is this uniqueness Lemma that ensures our main MACRS algorithm FPT\kl{KL: because}. 

%------------------------------------------------
%alg 3 is correct

\begin{restatable}{lemma}{subnetmaker}
    \label{lem:subnet-maker}
    %Algorithm~\ref{alg:make-rt-subnet}($\N$, $F$) returns a subnetwork $\N'$ such that $\N' \simeq \N^*$\comment{KL}{or strong iso? because labels should be preserved?} for $\N^*$ reticulation trimmed subnetwork of $\N$ with respect to $F \subset E_R(\N)$ and runs in $O(|E_R(\N)|)$ time. 

    Algorithm~\ref{alg:make-rt-subnet} on ($\N$, $F$) returns the reticulation-trimmed subnetwork $\N'$ of $\N$ with respect to $F$ if it exists, and NULL if not, and runs in time $O(|V(\N)|)$.

\end{restatable}
For proof of Lemma~\ref{lem:subnet-maker}, see Appendix.

\begin{restatable}{theorem}{reducedset}
    \label{thm:reduced-set}
    Algorithm~\ref{alg:reduced-set} correctly enumerates all reticulation-trimmed subnetworks of a network $\N$, and runs in time $O(3^{|R(\N)| }|V(\N)|)$. 

\end{restatable}

%\ml{[ML: I assumed here that level-1 networks have a linear number of edges.  I'm not sure, please check.]}
\begin{proof}

    It is already proved (Lemma~\ref{lem:disjoint-F}) that non-disjoint $F$ does not admit a reticulation-trimmed subnetwork, so it is correct to filter those. The remaining correctness follows from the exhaustive nature of the construction of all $F$ and by the correctness of Algorithm~\ref{alg:make-rt-subnet}
    
    As for the time complexity, filtering non-disjoint $F$ implies a threefold choice on each reticulation (we either include one, or none of its incoming edges, but not both by disjointness). Thus the size of the set is $O(3^{|R(\N)|})$.  Recalling that Algorithm~\ref{alg:make-rt-subnet} 
    can be implemented in time $O(|V(\N)|)$,  
    the total runtime for Algorithm~\ref{alg:reduced-set} is in $O(3^{|R(\N)|}|V(\N)|)$.
    
\end{proof}
    \subsection{An algorithm for \macrssimple}
        \label{ss:simple}
        %tex master: sub
\label{ss:simple}
A dynamic programming algorithm that solves the \macrssimple~problem in cubic time is given and proved in this section.
%in \ref{ss:simple-alg} and proved in \ref{ss:simple-alg-proof}.

%\subsubsection{MACRS-SIMPLE problem definition}
%\label{ss:simple-prob-def}
%\input{lvl1/simple/prob-def}
%\subsubsection{A MACRS-SIMPLE Algorithm}
%\label{ss:simple-alg}
%Tex Master : simple < lvl1 < main
%
Assume we have networks $\N_1$ and $\N_2$ as input to the \macrssimple~problem.
We assume that we have computed the set of biconnected components of $\N_1$ and $\N_2$ in a preprocessing step, along with the bridge edges.  This can be done in time $O(|V(\N)|)$, see \cite{HTE73}.
%\sout{These bridge edges constitute the rows (for $\N_1$) and columns (for $\N_2$) of the table $M$ in the algorithm below.}
%\comment{ML}{I got a bit confused since it's not the edges that are used to index $M$, but the vertices.  I tried to define those more precisely, make sure you agree.}
Since the networks considered are level-1, each biconnected component $B$ contains exactly one vertex $u$ that has no in-neighbor in $B$, and exactly one vertex $r$ that has no out-neighbor in $B$.  If $B$ is trivial, then $u = r$, and otherwise $r$ is a reticulation vertex and there are two edge-disjoint paths from $u$ to $r$ in $B$\cite{huson2010phylogenetic}
%\ml{ ML: pretty sure there is a [REF] for this structure.}. 
We refer to these two paths as \emph{component paths}, 
%as described in Lemma~\ref{lem:2paths}.
%\ml{[ML: If we need a lemma, it should be stated above. 
% For the conference version, I'd say it's ok to not have the lemma (that will take space).  I'm fine with doing as everyone else by citing the book.  For the full version it should be there though.]}
%\ots{OTS:These lemmas are very far below right now. What is the best way to link them here?}. 
The vertex $u$ will be called the \emph{root} of $B$ and denoted $\rho(B)$, and $r$ will be called the \emph{bottom} of $B$.
We let $\B_1$ be the set of biconnected components of $\N_1$ and $\B_2$ be the set of biconnected components of $\N_2$.  Finally for $i \in \{1, 2\}$, we denote $\rho(\B_i) = \{\rho(B) : B \in \B_i\}$, i.e. the set of roots in $\B_i$.  %\comment{ML}{confusing to use $r$ for root and reticulation.  I suggest renaming $r(\N)$ to $\rho(\N)$.}

%In a preprocessing step, the bridges of $\N_1$ and $\N_2$ are identified, as they are all the roots of the independent subnetworks in a network(Lemma~\ref{cor:bridges-root}). 
%We also label whether bridge heads are on trivial or non-trivial components. 
Using dynamic programming, we construct a table $M$ whose rows are the roots in $\rho(\B_1)$ and whose columns are the roots in $\rho(\B_2)$.
For $u \in \rho(\B_1), v \in \rho(\B_2)$, we define $\N_u$ as the subnetwork of $\N_1$ rooted at $u$, and $\N_v$ as the subnetwork of $\N_2$ rooted at $v$.
We then define $M[u, v]$ as the number of \emph{leaves} in a \macrssimple~of $\N_u$ and $\N_v$.
If $u$ is a tree vertex, its children are denoted $u_1$ and $u_2$. 
%Let $r_1$, $r_2$ be $\rho(\N_1)$ and $\rho(\N_2)$ respectively.

%\ots{OTS: this doesn't seem to be used; the algo defines $r_1$ and $r_2$ as reticulations}
%Let $r_1$, $r_2$ be the heads of the edges leaving $\rho(\N_1)$, $\rho(\N_2)$ respectively \comment{ML}{make sure this makes sense wrt final def of root that we choose}.

\begin{figure}
\centering
\resizebox{.5\linewidth}{!}{
\begin{tikzpicture}
    \coordinate (rho) at (0,0);
    %\path (rho) +(90:0.5) coordinate (root);
    %\draw[decorate,decoration={snake,amplitude=0.7mm}] (rho) -- (root);
    \coordinate (pl1) at (-1,-2);
    \coordinate (r1) at (0,-4);
    \coordinate (pr1) at (1,-1);
    \coordinate (pr2) at (1,-3);
    \draw[line width=4pt,opacity=0.7,yellow] (rho) -- (pl1) -- (r1);
    \draw[line width=4pt,opacity=0.5,green] (r1) -- (pr2) -- (pr1) -- (rho);
    \draw (rho) -- (pl1) -- (r1) -- (pr2) -- (pr1) -- (rho);
    \node[tree,red,label={[xshift=5.5mm,yshift=-4.5mm]$p^1_{l,1}$}] at (pl1){};
    \node[ret,red,label={[xshift=5mm,yshift=-4.5mm]$r_1$}] at (r1){};
    \node[tree,red,label={[xshift=5.5mm,yshift=-4mm]$p^1_{r,1}$}] at (pr1){};
    \node[tree,red,label={[xshift=5.5mm,yshift=-4mm]$p^1_{r,2}$}] at (pr2){};
    \foreach \v/\x/\y in {a/-2/-4,b/0/-5,d/3/-2,c/2.5/-4} {
        \coordinate (\v) at (\x,\y);
        \node[decorate,decoration={snake,amplitude=0.3mm},draw,circle,inner sep=5mm,anchor=north,yshift=-0.5mm] at (\v){};   
    }
    \draw (pl1) -- (a);
    \draw (pr1) -- (d);
    \draw (pr2) -- (c);
    \draw (r1) -- (b);
    \node[tree,label={[xshift=-6mm,yshift=-3mm]$h(p^1_{l,1})$}] at (a){};
    \node[tree,label={[xshift=8mm,yshift=-3.5mm]$h(p^1_{r,2})$}] at (c){};
    \node[tree,label={[xshift=7mm,yshift=-3.5mm]$h(p^1_{r,1})$}] at (d){};
    \node[tree,red,label={[xshift=6mm,yshift=-3mm]$\rho(B_i)$}] at (rho){};
    %----------the above subnet
    \coordinate (root) at (0,1);
    \node[tree,label={[xshift=3mm,yshift=-3mm]$v$}] at (root){};
    \draw (rho) -- (root);
    \path (root) +(90:1) coordinate (above);
    %\draw[decorate,decoration={snake,amplitude=0.7mm}] (root) -- (above);
    \draw (root) -- (above);
    \path (root) +(-155:2) coordinate (n);
    \node[decorate,decoration={snake,amplitude=0.3mm},draw,circle,inner sep=5mm,anchor=north,yshift=-0.5mm] at (n){};
    \draw (root) -- (n);
    \node[tree] at (above){};
    \path (above) +(90:0.7) coordinate (aboveabove);
    \draw[decorate,decoration={snake,amplitude=0.7mm}] (above) -- (aboveabove);
    \path (above) +(-15:2) coordinate (n2);
    \draw (above) -- (n2);
    \node[decorate,decoration={snake,amplitude=0.3mm},draw,circle,inner sep=5mm,anchor=north,yshift=-0.5mm] at (n2){};
\end{tikzpicture}
}
\caption{
\label{fig:macrs-simple}
In this figure, tree vertices as filled circles and reticulations as filled squares. A subnetwork is represented by a large open blob. vertices in red are in the same non-trivial biconnected component. Yellow edges are path $\pi^1_1$ and green edges are path $\pi^1_r$. Tree vertex $v$ is a trivial biconnected component itself such that $R(v) \neq \emptyset$. 
}
\end{figure}
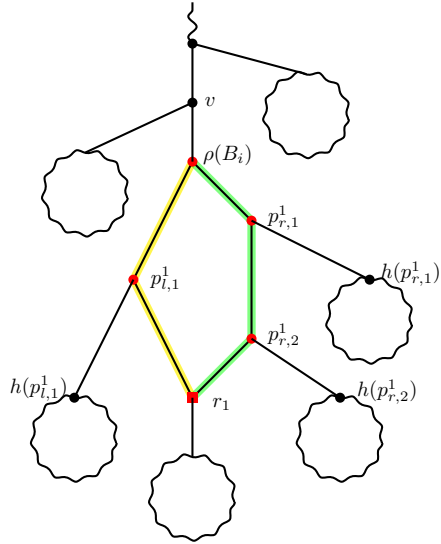

In $\N_1$, we denote the two component paths on the same non-trivial biconnected component by $\pi^1_l = p^1_{l,1}\ldots$ and $\pi^1_r = p^1_{r,1}\ldots$ and in $\N_2$ these paths will be denoted $\pi^2_l = p^2_{l,1}\ldots$ and $\pi^2_r = p^2_{r,1}\ldots$. For a vertex $p_i$ on path $\pi = p_1 \ldots$, let $h(p_i)$ be the child vertex of $p_i$ such that $h(p_i)\neq p_{i+1}$. In other words, the edge $(p_i,h(p_i))$ is a bridge pendant $\pi$ leading to a different biconnected component where $h(p_i)$ is rooting a distinct subnetwork. See Figure~\ref{fig:macrs-simple} for an illustration of the component paths and the described labelings for an example $\N_1$ network. 

We use Algorithm~\ref{alg:macrs-simple} to compute $M[u, v]$ for each $u \in \rho(\B_1), v \in \rho(\B_2)$ in postorder. We seek the result $M[\rho(\N_1), \rho(\N_2)] + |R(\N_1)|$ as $M$ records only the number of leaves in an \macrssimple~of $\N_1$ and $\N_2$. From this information we can calculate more about the general size of the network because they are binary, $ |V(\N)| = 2|L(\N)| + 2|R(\N)| -1  $.
Luckily, the number of reticulations in the solution is known ahead of time since it must have the same number of reticulations as each of the inputs. Note that we can also reconstruct the network that corresponds to the optimal size of the \macrssimple~of $\N_1$ and $\N_2$ by performing a traceback in the dynamic programming table.

%\ml{ML: the algorithm only handles a specific $u, v$.  To be complete, it requires two more loops, ``for $u \in \rho(\B_1)$ in postorder'' then ``for $v \in \rho(\B_2)$ in postorder''.  If we add those to the algorithm, it will probably mess up the indentation and the clarity.
%I suggest we present the algorithm in text blocks, instead of in an algorithm box.  We can say textually that we compute $M[u, v]$ for each $u \in \rho(\B_1), v \in \rho(\B_2)$ in postorder.  Then we can present each subcase in a different text block.  This will also have the advantage of not requiring a box that fits in one page.}
%\ots{Mention in text: 1. call MACRS-SIMPLE algo on u,v in postorder. 2. The final answer is the entry in the table for $\rho(\N_1)$ and $\rho(\N_2)$ 3. Mention that you need to add the number of reticulations (algo does not calculate that, but you know from the input networks) 4. Mention that you can get the number of vertices from the number of leaves and retics. 5. Mention that the algo only returns a value that is not $-\infty$ when the number of retics is the same in both input nets, we don't have to even call it when we know they don't have the same number of retics (optimization). }

%\ml{ML: figure would be nice to illustrate all the paths and $h(p_i)$'s.}

%Let the set $X_{11} = X(u_1) \cap X(v_1)$, $X_{22} = X(u_2) \cap X(v_2)$, $X_{12} = X(u_1) \cap X(v_2)$, and $X_{21} = X(u_2) \cap X(v_1)$

\begin{singlespace}
\begin{algorithm}[htpb]
\caption{}\label{alg:macrs-simple}
\hspace*{\algorithmicindent}\textbf{Input:} Two multi-networks $\N_1$, $\N_2$, vertices $u \in \rho(\B_1), v \in \rho(\B_2)$ \\
\hspace*{\algorithmicindent}\textbf{Output:} $M[u, v]$\\
%$|L(\N^*)|+|R(\N^*)|$, where $\N^*$ is a \macrssimple~of $\N_1$ and $\N_2$, or $-\infty$ if there is no solution\ots{OTS: Should we present the output as a table, with every cell representing the MACRS-SIMPLE of the two corresponding nodes?}

%\ml{[ML: someone will have to find a solution to $M_1, M_2$ going out of margins.  I don't wanna do it! (I did it last paper)]}

\begin{algorithmic}[1]
    
    \If{both $u$ and $v$ are trivial components}
        \If{$u$ or $v$ is a leaf} 
            \State 
                \begin{equation*}
                    M[u,v] =
                    \begin{cases*}
                        1 & if $ X(u)\cap X(v) \neq \emptyset $ and $R(u)\cup R(v)=\emptyset$ \\
                        -\infty & otherwise
                    \end{cases*}
                \end{equation*}
        \Else 
            \State for each $i \in \{1, 2\}, j \in \{1, 2\}$, define $X_{ij} = X(u_i) \cap X(v_j)$
            \State        $M[u,v]  = max(M_1,M_2)$ where
                    \begin{equation*}
                        M_1 =
                        \begin{cases*}
                          1 & if $X_{11} \neq \emptyset$ and $X_{22} = \emptyset$ and $R(u)\cup R(v)=\emptyset$ \\
                          1 & if $X_{11} = \emptyset$ and $X_{22} \neq \emptyset$ and $R(u)\cup R(v)=\emptyset$ \\
                          M[u_1, v_1]+ M[u_2, v_2] & if $X_{11} \neq \emptyset$ and $X_{22} \neq \emptyset$ \\
                          -\infty & otherwise 
                        \end{cases*}
                    \end{equation*}
                    \begin{equation*}
                        M_2 =
                        \begin{cases*}
                          1 & if $X_{12} \neq \emptyset$ and $X_{21} = \emptyset$ and $R(u)\cup R(v)=\emptyset$\\
                          1 & if $X_{12} = \emptyset$ and $X_{21} \neq \emptyset$ and $R(u)\cup R(v)=\emptyset$\\
                          M[u_1, v_2]+ M[u_2, v_1] & if $X_{12} \neq \emptyset$ and $X_{21} \neq \emptyset$ \\
                          -\infty & otherwise 
                        \end{cases*}
                    \end{equation*} 
            \EndIf
    \ElsIf{$u$ is a trivial biconnected component and $v$ is in a non-trivial biconnected component (or vice versa)}
        \State $M[u,v]= -\infty$
    \Else{ $u$ and $v$ are in non-trivial components with reticulations $r_1$, $r_2$ respectively and complement paths $\pi^1_l$, $\pi^1_r$, $\pi^2_l$, $\pi^2_r$}\label{algline:macrs-simple:pathstart}
        \State $M_1 = -\infty$
        \State $M_2 = -\infty$
        \If{$|\pi^1_l| = |\pi^2_l|$ and $|\pi^1_r| = |\pi^2_r|$}  
            \State
                \begin{equation*} 
                    M_1 = M[r_1, r_2] + \sum_{i=1}^{i=|\pi^1_l|}{M[h(p^1_{l,i}), h(p^2_{l,i})]} + \sum_{i=1}^{i=|\pi^1_r|}{M[h(p^1_{r,i}), h(p^2_{r,i})]}
                \end{equation*} 
        \EndIf
        \If{$|\pi^1_l| = |\pi^2_r|$ and $|\pi^1_r| = |\pi^2_l|$}
            \State 
                \begin{equation*} 
                     M_2 =   M[r_1,r_2] + \sum_{i=1}^{i=|\pi^1_l|}{M[h(p^1_{l,i}), h(p^2_{r,i})]} + \sum_{i=1}^{i=|\pi^1_r|}{M[h(p^1_{r,i}), h(p^2_{l,i})]}
                 \end{equation*}
        \EndIf
        \State $M[u, v] = \max(M_1, M_2)$
    \EndIf
    % \Else{ $u$ and $v$ are in non-trivial components with reticulations $r_1$, $r_2$ respectively and forbidden paths $\pi^1_l$, $\pi^1_r$, $\pi^2_l$, $\pi^2_r$}\label{algline:macrs-simple:pathstart}
    %     \If{$|\pi^1_l|=|\pi^1_r|=|\pi^2_l|=|\pi^2_r|$}
    %         \State let $M[u,v] =max(M[u,v] \text{ as defined by } $Eq~\ref{eq:m1}, $M[u,v] \text{ as defined by }$Eq~\ref{eq:m2})
    %     \ElsIf{$|\pi^1_l| = |\pi^2_l|$ and $|\pi^1_r| = |\pi^2_r|$}  
    %         \State
    %             \begin{equation} \label{eq:m1}
    %                 M[u,v] = M[r_1,r_2] + \sum_{i=1}^{i=|\pi^1_l|}{M[h(p^1_{l,i}), h(p^2_{l,i})]} + \sum_{i=1}^{i=|\pi^1_r|}{M[h(p^1_{r,i}), h(p^2_{r,i})]}
    %             \end{equation} 
    %     \ElsIf{$|\pi^1_l| = |\pi^2_r|$ and $|\pi^1_r| = |\pi^2_l|$}
    %         \State 
    %             \begin{equation} \label{eq:m2}
    %                  M[u,v]  =   M[r_1,r_2] + \sum_{i=1}^{i=|\pi^1_l|}{M[h(p^1_{l,i}), h(p^2_{r,i})]} + \sum_{i=1}^{i=|\pi^1_r|}{M[h(p^1_{r,i}), h(p^2_{l,i})]}
    %              \end{equation}
    %     \Else
    %         \State $M[u,v]= -\infty$%r1 and r2 attempting mapping but forbidden paths were of different lengths
    %     \EndIf
    % \EndIf
    %\State return $M[\rho_1,\rho_2] + |R(\N_1)|$\ml{ML: moved this to the end}
\end{algorithmic}
\end{algorithm}
\end{singlespace}

\begin{restatable}{theorem}{macrssimpletime}
    \label{thm:simple-time}
    Algorithm~\ref{alg:macrs-simple} runs in time $O(|V(\N_1)||V(\N_2)|(|V(\N_1)|+|V(\N_2)|))$.

\end{restatable}

\begin{proof}

    The algorithm fills a table $M$, a table of maximum size $|V(\N_1)||V(\N_2)|$, thus if we can show each table entry is calculated in at most linear ($|V(\N_1)|+|V(\N_2)|$) time, then the algorithm is cubic as claimed. 

    The preprocessing step to determine and label biconnected components is linear as it requires a modified depth-first search~\cite{HTE73}. Then, the calculations being performed for lines 1 through 9 consist of finding and checking the labelled components (linear), and checking up to 12 set intersections (linear) of a vertex's descendants leaves (linear to find). Lines 10 and on perform a linear number of table lookups/calls. The paths themselves are also linear to find as they are simply the paths that leave each child of the rooting vertex of the biconnected component and end on the next reticulation, the length of which can also be calculated on a single pass. 
    Thus the claim holds. 
    
\end{proof}

\begin{restatable}{theorem}{Mcorrect}
    \label{thm:M-correct}
    The entry $M[\rho(\N_1), \rho(\N_2)]$ correctly contains $|L(\N^*)|$ for $\N^* = \macrssimple (\N_1,\N_2)$ if one exists, and $-\infty$ otherwise.  
    
    %the number of leaves between a \macrssimple~between $\N_1$ and $\N_2$, if one exists, and $-\infty$ otherwise.
\end{restatable}

See Appendix for proof of Theorem~\ref{thm:M-correct}.

\subsection{Complexity of Algorithm~\ref{alg:macrs}}
    %tex master: main
%

%\ml{[ML: I suggest making this a subsection]}

\begin{restatable}{theorem}{complexity}
Let $\N_1, \N_2$ be two networks, let $n = \max(|V(\N_1)|, |V(\N_2)|)$, and $r = |R(\N_1)| + |R(\N_2)|$.  Then the \macrs~problem can be solved in time 
$O(3^r n^3)$.
\end{restatable}

\begin{proof}
    By Theorem~\ref{thm:reduced-set}, Algorithm~\ref{alg:reduced-set} can enumerate all reticulation-trimmed subnetworks of $\N_1$ and $\N_2$ in total time $O(3^{|R(\N_1)|}n + 3^{|R(\N_2)|}n) = O(3^r n)$.  The number of pairs of such networks for which we compute a \macrssimple~is $O(3^r)$, each of which can be handled in time $O(n^3)$ by Theorem~\ref{thm:simple-time}.  The total running time is thus $O(3^r n + 3^r n^3) = O(3^r n^3)$. 
    %Then, for each such $F \in F_1 \times F_2$, we call Algorithm~\ref{alg:make-rt-subnet} ($O(n)$) and Algorithm~\ref{alg:macrs-simple} ($O(n^3)$).In all, we have that Algorithm~\ref{alg:macrs} runs in $O(3^r n^3)$

\end{proof}

% -----------------
\section{Conclusion and discussion}
    %tex master: main

In this paper, we presented the first exact algorithm to find an MACRS of two rooted binary level-1 networks. The proposed approach starts by enumerating all reticulation-trimmed subnetworks for both input networks, and then compares all the possible pairs produced for each input network using a dynamic programming algorithm for the \macrssimple~problem. The enumeration step presented here is currently exponential in the sum of reticulation numbers of both input networks, and the \macrssimple~algorithm takes cubic time in the maximum number of vertices contained in the input networks.

In addition to the benefit of being able to extract a common subnetwork structure of maximum size from two orchard networks, the proposed algorithm permits to find a measure of the amount of differences between them.
As shown in our previous work~\cite{CDIST22}, there is a direct correspondence between finding an MACRS (more specifically, its size) and calculating one of the three equivalent distances presented in that work. As such, the algorithm presented here provides a first method to calculate exactly these distances. This can be used in the future to compare this distance with other distances (such as the mixed distance) or to evaluate the accuracy of different heuristic approaches.

\subsection*{Future extensions}

There is an obvious optimization that can be applied to the approach presented in this work related to the enumeration of the reticulation-trimmed subnetworks. 
Since the \macrssimple~algorithm by definition does not remove reticulations, comparing two input reticulation-trimmed subnetworks that do not share the same reticulation number or topology (in the sense that no mapping of the components containing reticulations can be made) will result in no solution. An obvious improvement to the enumeration step is to compare the topological relationships of the reticulations in both input networks (which, in the case of level-1 networks, can be modelled by trees), find the largest common reticulation topology between them, and start enumerating from there by gradually removing all possible reticulations. 
%Its fairly easy to see that for any solution to the \macrssimple~problem will need to have an embedded network isomorphism in all vertices above any reticulation in the network . thus we can immediately return a null result on any inputs to macrs-simple that do not have the same relationship between the (unlabeled) reticulations. On level-$1$ networks reticulations have a laminar relationship which can be modelled with trees. We then simply checkf for isomorphism on these pairs of trees as a filtering step, decreasing the number of inputs to macrs-simple. 
While this strategy does not achieve any additional formal bounding, it may reduce greatly the number of reticulation-trimmed subnetwork pairs to consider on many real inputs (potentially bringing it down to a linear number of pairs).

Another interesting avenue of work is to generalize our algorithm to higher level networks. A brief overview of a possible strategy would be to extend the \macrssimple~dynamic programming to consider, for each pair of biconnected components, all possible isomorphisms, find the maximum value and then summing to it the values of the exterior nodes that are matched in the isomorphism.

Attaching leaves to a non-orchard network was used previously to extend an approach to solve the minimum hybridization problem on any rooted phylogenetic network~\cite{linz2019attaching}. Exploring if and how a similar idea could be employed to generalize our proposed algorithm to non-orchard networks should be considered. 

Finally, as mentioned earlier, the complexity of our method is exponential in the sum of the number of reticulations in both input networks because of the enumeration step. Ideally, we could find an approach for which the complexity would depend only on the level of the two input networks, which we leave as an open problem.

%It is also possible to count how many subsets of F will induce a reticulation-trimmed subnetwork for a given network [where is this in context].

%\ml{ML: I suggest adding a `future extensions' section, or something similar, where we could discuss
%    (1) how to extend our algorithm to higher level networks.  The idea of the DP extension would be to have a table $M[B_i, B_j]$ indexed by pairs of biconnected components.  The value of this $M$ entry could be obtained by taking the maximum over every possible isomorphism between the biconnected components, each for each of them, summing the $M$ values of the exterior nodes that are matched in the isomorphism.
%    (2) how to improve the complexity analysis of the enumeration part, by arguing that in some cases, the number of trimmed subnets is linear, and that in general the complexity of enumerating depends on how much dependency there is between the reticulations.
%    (3) our complexity is exponential in $|R(\N_1)| + |R(\N_2)|$ because of the enumeration part.  It would be better to have complexity that depends only on the level of $\N_1$ and $\N_2$, which we leave as an open problem.}
% -----------------
\bibliographystyle{splncs04}
\bibliography{main}

\newpage
\section*{Appendix}
\setcounter{section}{0} 
%tex master: main
%\begin{appendices}

\section{Proof of Lemma~\ref{lem:sub-through-trim}(page~\pageref{lem:sub-through-trim})}
\subthroughtrim*
\begin{proof}

Let $\N$ and $\N'$ be networks such that $\N' \crof \N$.  
We use induction on $|E(\N)|$ to prove a slightly stronger statement. 
We show that for any $F \subseteq E_R(\N)$ such that there exists a CS $S'$ satisfying $\cs{\N}{S'} = \N'$ that removes the set of reticulation edges $F$, there exists a reticulation-trimmed subnetwork $\N''$ of $\N$ with respect to $F$ and a CS $S$ such that $\cs{\N''}{S} = \N'$.

The base case is $|E(\N)| = 1$. In this case we have a singleton network on a root, a single leaf, and an edge between them. There are no cherries and only $F = \emptyset$ is possible, and so all $\N' \crof \N$ have $\N' = \N$ and $S = \emptyset$ for $\cs{\N}{S} = \N'$. So the claim is trivially true. 

For the induction step, assume that the claim holds for all networks whose number of edges is strictly smaller than $|E(\N)|$. 
%\N$ such that $|V(\N)| + |E(\N)| \leq k$ for $k > 3$.
%For the induction step, assume that $|V(\N)| + |E(\N)| > k$.

Let $F \subseteq E_R(\N)$, and suppose there is a CS $S'$ such that $\cs{\N}{S'} = \N'$, and that the set of reticulation edges removed by $S'$ is $F$. If every reduction in $S'$ is simple, then $F = \emptyset$ and $\N$ is itself a reticulation-trimmed subnetwork of $\N$ with respect to $F = \emptyset$, in which case the statement holds. Otherwise, 
%let $F$ be the set of reticulations edges removed by $S'$.  Moreover, 
let $(u, v) \in F$ be the first reticulation edge of $\N$ removed by a reduction in $S'$, say $(u,v)$ is removed by cherry $S'_i$. On network $\cs{\N}{S'_{(0:i)}}$, $u$ and $v$ must both have leaf children. This implies that there are no reticulations in $\below(u,\N)$. Thus, all cherries $(x, y)$ of $S_{(0:i)}$ such that $\{ x,y \} \in \below(u,\N)$ are simple. 

Assume for now that at least one such simple reduction exist, and let $S'_h = (x, y)$ be the first of them. With this, we see that $(x,y)$ is a cherry on $\N$ throughout every step of the reduction of $\N$ by $S'_{(0:h)}$, including on $\N$. We may therefore assume that $(x, y)$ is the first reduction of $S'$ by Theorem~\ref{thm:any-order}.
%[is that justified?  it's really needed].}  added lemma reference
Thus $\N'$ is obtained from $\cs{\N}{(x,y)}$ by applying the CS $S'_{(1:|S|)}$, which we know removes the set of reticulation edges $F$.  
By induction, there exists a reticulation-trimmed subnetwork $\N^*$ of $\cs{\N}{(x,y)}$ with respect to $F$
%\kl{KL: F is derived from a CS, does that imply it admits an r-t subnet? we don't prove that} \ml{ML: correct, we can't assume it is wrt $F$, it could be another $F'$, needs fixing} 
and a simple CS $S^*$ such that $\cs{\N^*}{S^*} = \N'$. 
Let $T$ be a smallest CS such that results in $\N^*$ after applying it on $\cs{\N}{(x,y)}$.  Then $\N^*$ can be obtained from $\N$ by applying the CS $(x,y) \cdot T$.
We note that reduction $(x, y)$ is required in any CS that results in a trimmed-reticulation subnetwork of $\N$ with respect to $F$ (since the cherry $(x, y)$ is below $u$ and $(u, v)$ needs to be removed) and it follows by the minimality of $T$ on $\cs{\N}{(x,y)}$ that $(x,y) \cdot T$ is minimum on $\N$.
% $S'^*$ that contains reticulated reductions if and only if they reduce an edge in $F$. In this way, we find CS $(x,y) \cdot S'^*$ such that $\cs{\N}{(x,y)\cdot S'^*} = \N^*$ and $\cs{\N^*}{(x,y)\cdot S^*} = \N'$ and is minimal and contains a reticulated reduction if and only if it removes an edge in $F$, 
Thus $\N^*$ is a reticulation-trimmed subnetwork of $\N$ with respect to $F$ and the claim holds. 

%By Theorem~\ref{thm:any-order}, we may assume that $(x, y)$ is the first element of $S''$ where $\cs{\N}{S''} = \N''$ and such that $S''$ is minimal and contains only . %Now, observe that in any reticulation-trimmed subnetwork of $\N$ with respect to $F$, the reduction $(x, y)$ is necessary to be able to remove $(u, v)$.  We can now assume that $(x, y)$ is applied first to obtain such a reticulation-trimmed subnetwork and it follows that $\N^*$ is also a reticulation-trimmed subnetwork of $\N$ with respect to $F$, %$\cs{\cs{\N}{(x,y)}}{S^*} = \N'$ %$\cs{\N^*}{S^*} = \N''$.
%[We may assume that $(x, y)$ is applied first to obtain such an r-t subnet](**), and it follows that $\N^*$ is also an r-t subnet of $\N$ wrt $F$.  Our statement holds because $\N^*[S^*] = \N''$.

It is also possible that $(x, y)$ does not exist as a simple reduction, which occurs when $u$ and $v$ each already have a leaf child in $\N$.  In this case, $(x,y)$ is a reticulated cherry on $\N$ such that $p(x) = v$ and $p(y) = u$. As in the previous case, we may assume that $(x, y)$ is the first reduction of $S'$.  Let $F' = F \setminus \{(u, v)\}$.
By induction there is a reticulation-trimmed subnetwork $\N^*$ of $\cs{\N}{(x,y)}$ with respect to $F'$ and there exists $S^*$ such that $\cs{\N^*}{S^*} = \N'$. We also have a CS $T$ that is minimum and contains a reticulated reduction if and only if it removes an edge in $F'$. 
As before, the reduction $(x, y)$ is required in any reticulation-trimmed subnetwork of $\N$ with respect to $F$, and it follows by the minimality of $T$ that $(x, y) \cdot T$ yields such a network.  Again, $\N^*$ is a reticulation-reduced subnetwork of $\N$ with respect to $F$ and there is the simple CS $S^*$ such that $\cs{\N^*}{S^*} = \N'$.
% We similarly see that $\cs{\N}{(v,u) \cdot S'^*} = \N^*$ and $(v,u) \cdot S'^*$ is minimal and contains a reticulated reduction if and only if it removes an edge in $F$ (as the removal of $(u,v)$ is required in this case and is done so by cherry $(v,u)$), so $\N^*$ is a reticulation-reduced subnetwork of $\N$ with respect to $F$ and there is the simple CS $S^*$ such that $\cs{\N^*}{S^*} = \N'$ so the claim holds. 

%[In this case, we may assume that $(v, u)$ is the first reduction of $S$](***).%Let $\N_{v, u}$ be the network obtained after applying $(v, u)$ on $\N$.  As in the previous case, by induction there is $\N^*$ such that [...].%Moreover, in any r-t subnet of $\N$ wrt $F$, $(v, u)$ is necessary.  [same as last paragraph]%(*) (**) (***) seem obvious, but need some argument, or maybe there is some lemma about ordering operations between independent parts of the networks.
\qed
\end{proof}

\section{Proofs from Section~\ref{ss:r-t-subnet-enum}: Lemma~\ref{lem:disjoint-F}(page~\pageref{lem:disjoint-F}), Lemma~\ref{lem:topsort}(page~\pageref{lem:topsort}), Lemma~\ref{lem:unique-rt-sub}(page~\pageref{lem:unique-rt-sub}), Lemma~\ref{lem:subnet-maker}(page~\pageref{lem:subnet-maker})}

\disjointF*
\begin{proof}
    Say $F$ is not disjoint, then there are two cases.
    First, say $F$ contains edges $(u,v)$ and $(w,v)$. The reduction on the network must first reduce one of these edges, say $(u,v)$. This reduction removes the vertex $v$, and thus the edge $(w,v)$ can no longer be in the network. In this way no single CS can reduce both the reticulation edges leading into the same reticulation.
    Next, assume $F$ is not disjoint because it contains edges $(u,v)$ and $(u,w)$. 
    This is not possible in a level-1 network, because reticulations $v, w$ would be contained in the same biconnected component.
    %The contradiction here is evident, the network $\N$ must be level-$1$. See Lemma~\ref{lem:2paths}.
    %\kl{maybe this needs to be somewhere else rather than in this proof?}
\end{proof}

\topsort*
\begin{proof}
    This claim is evidence by the topological partial ordering on $R(\N)$, which exists on any network. Then, note that $F$ is disjoint by Lemma~\ref{lem:disjoint-F}. 
   % \ml{[ML: is it possible to not rely on Corollary 1? I think it's ok to say that $R(\N)$ can be topologically ordered (which is true in any network, actually)]}
 %\kl{KL: so is this lemma needed at all? we just define topological sort, on the reticulations so its just implied in $E_R$ without this lemma}
\end{proof}

\onertsubnet*
\begin{proof}
    Let $\N$ be a network and $F \subseteq E_R(\N)$ be a set.

    We claim there are not two non-strongly isomorphic reticulation-trimmed subnetworks with respect to $F$. We proceed by induction on $|V(\N)|+|E(\N)|$.

    In the base case, $|V(\N)|+|E(\N)| \leq 3$ and $\N$ is a singleton network on one edge, with a root and a leaf as endpoints. In this case, $F = \emptyset$ and the reticulation-trimmed subnetwork of $\N$ with respect to $F$ is $\N$ and $\N = \N$. There are no cherries on $\N$ so there is not more than one reticulation-trimmed subnetwork of $\N$ with respect to $F$.

    Assume the claim holds for all networks with strictly less vertices and edges than $\N$. %$|V(\N)|+|E(\N)| \leq k$ for $k > 3$. 
    %In the inductive step, assume $|V(\N)|+|E(\N)| > k$. 
    If there is no reticulation-trimmed subnetwork of $\N$ with respect to $F$, then the claim holds. Assume such a network, $\N'$, exists. If $F = \emptyset$ then $\N$ is the only reticulation-trimmed subnetwork of $\N$ with respect to $F$, as making any cherry reductions would not be minimum (since $\cs{\N}{\emptyset} = \N$ is already sufficient to remove empty $F$). The claim holds in this case, so we now assume $F \neq \emptyset$.

    Let edge $e \in F$ be (arbitrarily) one of the lowest edge in the topological sort on $F$ (which exists, Lemma~\ref{lem:topsort}). Because $\N'$ exists, there is at least one cherry below an endpoint of $e$ ($e$ may be in the cherry itself). Let one such cherry be called $(x,y)$. 

    Next assume $(x,y)$ is such that $(p(y),p(x)) \neq e$.  We know that $(x, y)$ is not reticulated, as otherwise $e$ would not be one of the lowest reticulations in $F$.
    %is reticulated. No edge containing the endpoint $x$ is in $F$, otherwise $e$ would not be one of the lowest. 
    Therefore $(x,y)$ is not reticulated, and thus simple.  
    %So, $(x,y)$ such that $(p(y),p(x)) \neq e$ is a simple cherry. 
    Because $e$ is in $F$, it will have to be removed, and must have leaves under its endpoints to do so, thus $(x,y)$ needs to be reduced in any CS $S$ such that $\cs{\N}{S}$ is a reticulation-trimmed subnetwork of $\N$ with respect to $F$.
    Moreover, by Theorem~\ref{thm:any-order}, we may assume that any such CS $S$ starts with $(x, y)$ as otherwise $(x, y)$ can be removed first without affecting the resulting network.
    %There are many CSs that may yield $\N'$, are minimal and contain a reticulated reduction if and only if it removes an edge in $F$. By Theorem~\ref{thm:any-order}, we may assume that there is such a CSs $S$ that starts with cherry $(x,y)$.
    It follows that we may assume that, for every CS $S$ such that $\cs{\N}{S}$ is a reticulation-trimmed subnetwork of $\N$ with respect to $F$, applying the first reduction results in $\cs{\N}{(x, y)}$.  
    %Notice that applying $S_{(1:|S|)}$ on $\cs{\N}{(x, y)}$ yields $\N'$, which must also be a reticulation-trimmed subnetwork of $\cs{\N}{(x, y)}$ with respect to $F$ (in particular, $|S_{(1:|S|)}|$ is minimum as otherwise, $|S|$ would not be minimum).
    %Let $\N_{xy}'$ the reticulation-trimmed subnetwork, , of $\cs{\N}{(x,y)}$ with respect to $F$. \kl{$\N_{xy}'$ does exist...}. 
    Then applying $S_{(1:|S|)}$ on $\cs{\N}{(x, y)}$ must yield a reticulation-trimmed subnetwork of $\cs{\N}{(x, y)}$ with respect to $F$ (in particular, $|S_{(1:|S|)}|$ is minimum as otherwise, $|S|$ would not be minimum).
    By the induction hypothesis, $\cs{\N}{(x,y)}$ does not have two non-strongly isomorphic reticulation-trimmed subnetworks with respect to $F$.  
    Since we may assume that all CSs $S$ applicable to $\N$ that result in such a trimmed subnetwork go through $\cs{\N}{(x, y)}$, it follows that $\N$ also does not have two non-strongly isomorphic reticulation-trimmed subnetworks with respect to $F$. 
    % Therefore there exists a CS $S_{xy}'$ such that $\cs{\cs{\N}{(x,y)}}{S_{xy}'} = \N_{xy}'$ and such that $|S_{xy}'|$ is minimal and contains reticulated reductions if and only if it removes an edge in $F$. 
    
    %It is evident that $(x,y)$ must be removed in any reticulation-trimmed subnetwork of $\N$ with respect for $F$ since $(x,y)$ is simple and occurs below $e \in F$. 
    %Thus we have CS $(x,y)\cdot S_{xy}'$ such that $\cs{\N}{(x,y) \cdot S_{xy}'} = \N'$, and where $|S'_{xy}|$ is minimal and $(x,y)\cdot S_{xy}'$ contains a reticulation reduction if and only if it removes an edge in $F$. By definition of cherry reductions, every $\cs{\N}{(x,y)}$ is strongly isomorphic, and with the induction hypothesis, the claim holds. 

    Finally, assume that $(x,y)$ is reticulated and $(p(y),p(x)) = e$. 
    %This cherry must exist since we assumed $\N'$ exists, thus $e$ must be removable. 
    As in the previous case, note that $(x, y)$ must be present in any CS $S$ such that $\cs{\N}{S}$ is a reticulation-trimmed subnetwork of $\N$ with respect to $F$, and we may further assume that any such CS starts with $(x, y)$.
    Then, any such CS first goes through $\cs{\N}{(x, y)}$, and then by minimality, results in a reticulation-trimmed subnetwork of $\cs{\N}{(x, y)}$ with respect to $F \setminus \{(p(y), p(x))\}$.  By induction, there is only one such network, and thus also only one reticulation-trimmed subnetwork of $\N$ with respect to $F$. 
    %There are many CSs that may lead to $\N'$, and by Theorem~\ref{thm:any-order}, any of them may start with $(x,y)$. 
    % Thus we have network $\cs{\N}{(x,y)}$, which by the induction hypothesis has a reticulation-trimmed subnetwork $\N_{xy}''$ of $\N$ with respect to $F \setminus e$. Therefore, there exists CS $S_{xy}''$ such that $|S_{xy}''|$ is minimal, and it contains a reticulated reduction if and only if it removes an edge in $F$. Then, we also find the sequence $(x,y) \cdot S_{xy}''$ which, when applied to $\N$, contains a reticulated cherry if and only if it removes an edge in $F$, and for which $|(x,y) \cdot S_{xy}''|$ is minimal, thus $\N_{xy}''$ is a reticulation trimmed subnetwork of $\N$ with respect to $F$ and the claim holds. 

   % \kl{In the case there is no r-t subnet of $\cs{\N}{(x,y)}$ then... .}
    
    %Assume there is $\N'$, a reticulation-trimmed subnetwork of $\cs{\N}{(x,y)}$ with respect to $F\setminus e$.

  %\kl{There are many CSs that yield $\N'$.  ... we may assume that each such CS can start with $(x,y)$ (which is under some (lowest?) e in F). argue its simple. also there is the case where there are no simple cherries on the network. in which case would have to argue whether an edge from that cherry is in F or not. Next, assume $(x,y)$ is simple.}
  \qed
\end{proof}

\subnetmaker*
\begin{proof}

    First, there is always a topological sort on $F$ (Lemma~\ref{lem:topsort}) and $F$ is disjoint (Lemma~\ref{lem:disjoint-F}).
    
    %because $\N$ is level-$1$, there is always a partial order of the reticulations, and therefore of any set of disjoint reticulation edges. See Lemma~\ref{cor:bridges-root}. \ml{ML: ae need that there is no $(a, b) (c, b)$ in $F$} \kl{KL: meaning you can reach the first one in the order before the next one.. i.e. its a partial order}

    If a network is not returned, then it must be that $R(u) \cup R(v) \setminus \{ v \} \neq \emptyset$ on one of the (possibly) partially reduced subnetwork for some edge $(u,v)\in F$. In this case, $F$ does not admit a reticulation-trimmed subnetwork since there is no edge in $F$ that corresponds the reduction of all reticulations below $u$, and $v$, a requirement for the reduction of reticulation $v$. %(See lemma \comment{KL}{same as above})
    
    Assume a network is returned, we claim the algorithm is correct in this case.

    We will show this claim by constructing a CS $S$ such that $\cs{\N}{S} = \N'$, and by counting the exact number of reticulation cherries which we show will remove the desired $F$. Finally, we will show $S$ is minimum. 

    Say $F'$ has the order $\langle (u_1,v_1)(u_2,v_2)\ldots \rangle$. For each $(u_i,v_i)$, in order, we construct a CS $S_i$ on the partially reduced $\N$, then let $S = S_1 \cdot S_2 \ldots$. Note that this means we construct each $S_i$ on $\cs{\N}{S_1\cdot S_2 \cdot \ldots \cdot S_{i-1}}$. There are distinct subnetworks rooted on
    %they are rooted below u, v not on them....
    $u_i$ and $v_i$ so there are CSs $S^u_i$, $S^v_i$ that are complete for each respective subnetwork. Note how, since we assume a network was returned, $R(u_i) \cup R(v_i) \setminus \{ v_i \} = \emptyset$, so $S^u_i$ and $S^v_i$ are simple. Say they reduce the subnetworks to leaves $l^u_i$ and $l^v_i$ respectively (so that $p(l^u_i) = u$ and $p(l^v_i) = v$). Let $S_i = S^u_i \cdot S^v_i \cdot (l^v_i, l^u_i)$. The cherry $(l^v_i, l^u_i)$ will reduce the reticulation edge $(u,v)$ under these conditions. 

    In this way, there is a reticulation cherry in $S$ if and only if it reduces an edge in $F$.  Furthermore, $S$ is minimum since we construct $S_i$ on the network $\cs{\N}{S_1 \cdot S_2 \cdot \ldots \cdot S_{i-1}}$ and we have selected only the cherry reductions that are necessary and sufficient for the reduction of the targeted reticulation edges. 

    The claimed running time is straightforward.  We can obtain a topological sort on $F$ using a standard topological sort of $\N$ obtained in time $O(|V(\N)| + |E(\N)|) = O(|V(\N)|)$ (since our networks are binary).  Then the algorithm only iterates over $F$ and replaces subnetworks by multi-labeled leaves, which can be handled in time $O(|V(\N)|)$.
    %Since $F$ is not disjoint, $|F| \leq |R(\N)|$, the construction in lines 5-8 is at most linear, so the total runtime is in $O(|R(\N)|)$. 

\end{proof}

\section{Proof of Theorem~\ref{thm:M-correct}, page~\pageref{thm:M-correct}}

Lemma~\ref{lem:forbidden} and Corollary~\ref{cor:forbidden-isomorphism} is required to justify the arguments in the proof for Theorem~\ref{thm:M-correct}.

\begin{restatable}{lemma}{forbidden}
\label{lem:forbidden}

  Let $\N$ be a network and $S$ be any CS on $\N$. If, for $r \in R(\N)$, $r \in \cs{\N}{S}$, then $v\in \above(r, \N) \in \cs{\N}{S}$.

\end{restatable}
\begin{proof}
    
    Assume $\exists r \in \N, \in \cs{\N}{S}$, $v\in \above(r,\N), \notin \cs{\N}{S}$. First, $v$ is not a leaf as a leaf cannot be above any other vertex. There must be some cherry in $S$, say $S_i=(x,y)$ such that $v=p(x)$, $v=p(y)$ or $v=p(x)=p(y)$ in $\cs{\N}{S_{[(0:i)}}$. Since $r\in \cs{\N}{S}$, we have that $r\in \cs{\N}{S_{(0:i)}}$ and since $v\in\above(r,\N)$ we have that $v\in\above(r,\cs{\N}{S_{(0:i)}})$ thus the only orientation we can have is $v=p(y)$, $r=p(x)$ in the reticulated cherry $(x,y) \in \cs{\N}{S_{(0:i)}}$. But $x$, $y$, $p(x)$, $p(y)$ are removed in $\cs{\N}{S_{(0:i]}}$ contradicting that $r \in \cs{\N}{S}$.

\end{proof}

Note that Lemma~\ref{lem:forbidden} implies that after any simple reduction by a CS $S$ on a network $\N$, that since all $r\in R(\N) \in \cs{\N}{S}$ then $\bigcup_{r\in R(\N)}\above(r,\N) \in \cs{\N}{S}$. Noting this, it follows that

\begin{cor}
    \label{cor:forbidden-isomorphism}
    $\forall$ networks $\N_1$, $\N_2$ and $\forall \N^*= MACRS-SIMPLE(\N_1,\N_2)$ such that $\N^*$ is non-null,
    $\exists$ and edge-preserving bijective function $$f:\bigcup_{r\in R(\N_1)} \above(r,\N_1)\rightarrow \bigcup_{r\in R(\N_2)} \above(r,\N_2)$$.
\end{cor}

\Mcorrect*
\begin{proof}

    Given input networks $\N_1$ and $\N_2$, 
    %for any head of a bridge $u \in \N_1, v \in \N_2$,
    for any $u \in \rho(\B_1), v \in \rho(\B_2)$,
    let the subnetwork of $\N_1$ rooted on $u$ be called $\N_u$ and let the subnetwork of $\N_2$ rooted on $v$ be called $\N_v$. We claim that $M[u,v]$ as we defined it always contains the number of leaves of a \macrssimple~between $\N_u$ and $\N_v$.  In particular, this will show our desired result with $u = \rho(\N_1), v = \rho(\N_2)$.
    The proof is by induction on $|V(\N_u)|+|V(\N_v)|$. 

    %\medskip
    %\noindent 
    %\textbf{Base Case}

    %\noindent 
    As a base case, suppose that $|V(\N_u)| = 1$ and $|V(\N_v)| = 1$, i.e. $u$ and $v$ are both leaves. 
        If $X(u)\cap X(v)\neq \emptyset$, we put $M[u, v] = 1$, which is correct since $\N_u$ and $\N_v$ are weakly isomorphic.  
        If $X(u)\cap X(v) = \emptyset$, we put $M[u,v]= -\infty$, which is correct since there is \macrssimple~between networks that do not share a leaf label.

    %\medskip
 
    %\noindent 
    %\textbf{Inductive Step}

    %\noindent 
    Let us now consider the inductive step.
        For the rest of the proof, for any $u' \in V(\N_1), v' \in V(\N_2)$, we denote by $\N^*_{u', v'}$ a \macrssimple~between $\N_{u'}$ and $\N_{v'}$, if one exists (otherwise, $\N^*_{u', v'}$ is undefined).
        As an inductive hypothesis, we assume that $M[u',v']$ is the number of leaves in $\N^*_{u', v'}$, for any $u', v'$ such that $|V(\N_{u'})| + |V(\N_{v'})| < |V(\N_u)|+|V(\N_v)|$.  
        
        The proof is split into cases.

    \medskip 
    \noindent 
    \textit{Case: $u$ and $v$ are trivial, one is a leaf.} If $X(u)\cap X(v) \neq \emptyset$ and $R(u)\cup R(v) = \emptyset$ then $M[u,v]=1$. This is correct since a complete reduction can proceed on both networks without any reticulations and with at least one leaf in common, a requirement for any network to be isomorphic with a singleton network.   Moreover, there cannot be more than one leaf since $u$ or $v$ is itself a leaf.  For the same reason, when $X(u)\cap X(v) = \emptyset$, $M[u, v] = -\infty$ is obviously correct.  When $R(u)\cup R(v) \neq \emptyset$, $M[u, v]$ is correct because a \macrssimple~of $u$ and $v$ can only be a leaf, but this cannot be achieved since one of the networks has an unremovable reticulation.

    \medskip
    \noindent 
    \textit{Case: $u$ and $v$ are trivial and $R(u) \cup R(v) = \emptyset$, $X(u_1)\cap X(v_1)\neq \emptyset$ and $X(u_2)\cap X(v_2) = \emptyset$ or $X(u_1)\cap X(v_1) = \emptyset$ and $X(u_2)\cap X(v_2) \neq \emptyset$ or $X(u_1)\cap X(v_2)\neq \emptyset$ and $X(u_2)\cap X(v_1) = \emptyset$ or $X(u_1)\cap X(v_2) = \emptyset$ and $X(u_2)\cap X(v_1) \neq \emptyset$.} In this case, $M[u,v]=1$ by line 6. Neither $u$ nor $v$ have reticulations below them, thus any reduction may proceed. In fact, the reduction on $\N_u$ and $\N_v$ must be complete to obtain an isomorphic network since there is no shared leaf below one child of $u$ and one child of $v$, thus that child must be removed to reach any MACRS-SIMPLE($\N_u$, $\N_v$), requiring a cherry on $u$ and $v$. Luckily, there is a leaf shared below one child of $u$ and one child of $v$ and so a singleton isomorphic network is possible, so this case is correct. 

    \medskip 
    \noindent 
    \textit{Case: $u$ and $v$ are trivial, $R(u) \cup R(v) \neq \emptyset$, and $X(u_1)\cap X(v_1) = \emptyset$ or $X(u_2)\cap X(v_2) = \emptyset$ or $X(u_1) \cap X(v_2) = \emptyset$ or $X(u_2)\cap X(v_1) = \emptyset$.} In this case line 2 resolves to true so we calculate $M[u,v]$ on line 6. We find that $M[u,v] = -\infty$ since $R(u)\cup R(v) \neq \emptyset$. A complete reduction is required to reach the required isomorphic singleton in this condition, but the presence of a reticulation prevents this. 

    \medskip 
    \noindent 
    \textit{Case: $u$ is trivial and $v$ is not trivial or $v$ is trivial and $u$ is not trivial.} In this case we always resolve $M[u,v]= -\infty$ by line 8 and 9. This is indeed the correct case, the presence of a reticulation in $\N_v$ and not in $\N_u$ (or in $\N_u$ and not $\N_v$) makes an isomorphic network unreachable by simple reductions alone.

    \medskip 
    \noindent 
        \textit{Case: $u$ and $v$ are trivial, both are not leaves, and $X(u_1)\cap X(v_1)\neq \emptyset$ and $X(u_2)\cap X(v_2) \neq \emptyset$ or $X(u_1)\cap X(v_2)\neq \emptyset$ and $X(u_2)\cap X(v_1) \neq \emptyset$}. 
        In this case we calculate $M[u,v]$ on line 6. Regardless of any reticulations that may be below $u$ or $v$, we put $M[u,v]$ as the maximum between $M[u_1,v_1] + M[u_2,v_2]$ and $M[u,v] = M[u_1,v_2] + M[u_2,v_1] $. We can assume, by the inductive hypothesis, that $M[u_1,v_1]= |L(\N^*_{u_1,v_1})|$ and $M[u_2,v_2] = |L(\N^*_{u_2,v_2})|$ (likewise $M[u_1,v_2]= | L(\N^*_{u_1,v_2})|$ and $M[u_2,v_1]=|L(\N^*_{u_2,v_1})|$).
        %since $|V(\N_{u_1})|+ |V(\N_{v_1})| \leq k$ and $|V(\N_{u_2})|+ |V(\N_{v_2})| \leq k$ (likewise $|V(\N_{u_1})|+ |V(\N_{v_2})|\leq k$ and $|V(\N_{u_2})|+ |V(\N_{v_1})| \leq k$). 
        It is not difficult to see that if $\N^*_{u, v}$, a \macrssimple~of $\N_u$ and $\N_v$, exists, then it can be obtained by joining a \macrssimple~of $\N_{u_1}, \N_{v_1}$ with a \macrssimple~of $\N_{u_2}, \N_{v_2}$ under a common parent, or by joining a \macrssimple~of $\N_{u_1}, \N_{v_2}$ with a \macrssimple~of $\N_{u_2}, \N_{v_1}$ under a common parent.  In the current case, $M_1 = M[u_1, v_1] + M[u_2, v_2]$ and $M_2 = M[u_1, v_2] + M[u_2, v_1]$ correspond to constructing these two possible networks, and since they contain the correct values by induction, $M = \max(M_1, M_2)$ is correct.
        % The summation is correct, there exists a \macrssimple~between $\N_u$ and $\N_v$ consisting of joining the \macrssimple~that pairs the children of $u$ with the children of $v$\ml{ML: notion of pairing is informal}. This is evident by Corollary~\ref{cor:bridges-root} which demonstrates how any edges leaving a biconnected component (like the direct children of a trivial component) are themselves roots of disjoint networks. In fact, there is no other construction of a network that does not pair up children networks as described\kl{KL: why not?}, and we find the maximal solution since we have selected the maximum among these possible networks.

        \medskip 
        \noindent 
        \textit{Case: $u$ and $v$ are both non-trivial  
        and $M[r_1,r_2]\neq  -\infty$ for reticulation $r_1$, $r_2$ in $u$'s, $v$'s biconnected components respectively, and $M[h(p^1_i),h(p^2_i)] \neq -\infty$ for all $i$ in any $p^1\in \pi^1_l, \pi^1_r$ or $p^2\in \pi^2_l,\pi^2_r$)}. In this case we resolve  $M[u,v] = M[r_1,r_2] + \sum_{i=1}^{i=|\pi^1_l|}{M[h(p^1_{l,i}), h(p^2_{l,i})]} + \sum_{i=1}^{i=|\pi^1_r|}{M[h(p^1_{r,i}), h(p^2_{r,i})]}$ or $M[u,v]  =   M[r_1,r_2] + \sum_{i=1}^{i=|\pi^1_l|}{M[h(p^1_{l,i}), h(p^2_{r,i})]} + \sum_{i=1}^{i=|\pi^1_r|}{M[h(p^1_{r,i}), h(p^2_{l,i})]}$. %These cases are symmetric so for simplicity we will assume $M[u,v]$ has resolved to the first case.

        Each table reference in this summation returns a value that is correct for that subnetwork, by the induction hypothesis, as every subnetwork $\tilde{\N}$ of $\N_u$ ($\neq \N_u$) and $\N_v$ ($\neq \N_v$) is smaller. The summation itself is also correct. This is evident by noting that the biconnected components on $u$ and $v$ only contain vertices along $\pi_l$ and $\pi_r$\cite{huson2010phylogenetic}, so all vertices in the components are accounted for. Furthermore, all bridges must lead to disjoint networks.
        %Corollary~\ref{cor:bridges-root}, all edges leaving the component paths are themselves bridges to disjoint networks. 
        Finally, by Corollary~\ref{cor:forbidden-isomorphism} the only possible networks are constructed by joining up child networks 
        %\kl{KL: does this COR actually say that? it is the COR that insists there is a bijection between "forbidden" vertices} 
        that pair vertices in the order evident by the forbidden paths and their independent subnetwork children/siblings. Since we consider the maximum among all such possible networks, the solution is maximal. It is for this same reason when $u$ and $v$ are non-trivial but the conditions are such that $M[u,v]= - \infty$ by line 18, or by an operand being $-\infty$ in line 14 or line 16, that $M[u,v] = -\infty$ is correctly found.    
         
\end{proof}

\end{document}